\documentclass[11pt]{article}
\usepackage{color}
\usepackage{emlines}
\usepackage{array}
\usepackage{graphics}
\usepackage{graphicx}
\usepackage{epsfig}
\usepackage{amsmath}
\usepackage{amssymb}

\makeatletter
\newcommand{\doublespacing}{\let\CS=\@currsize\renewcommand{\baselinestretch}{1.25}\tiny\CS}
\newtheorem{theorem}{Theorem}
\newtheorem{lemma}{Lemma}

\newcommand {\tab} {\hspace*{2em}}
\newcommand {\ccc} {\textcolor{black}} 

\begin{document}

\date{}
\title {\Large {\bf
Scheduling algorithm to select $k$ optimal programme slots in television channels: A graph theoretic approach}}

\author{Madhumangal Pal$^*$ and Anita Pal$^\dag$
 \\
 $^*$Department of Applied
Mathematics with Oceanology and
  Computer Programming,
  \\Vidyasagar University, Midnapore -- 721 102, India.\\
{\sf e-mail:mmpalvu@gmail.com; Mobile no. (+91) 9932218937}\\
$^\dag$Department of Mathematics\\
 National Institute of Technology Durgapur\\
 Mahatma Gandhi Avenue, Durgapur-713 209\\
 West Bengal, India\\
{\sf  email: anita.buie@gmail.com}\\
}
\maketitle

\subsection* {\centering Abstract }
In this paper, it is shown that all programmes of all television
channels can be modelled as an interval graph. The programme slots
are taken as the vertices of the graph and if the time duration of
two {programme slots} have non-empty intersection, the
corresponding vertices are considered to be connected by an edge.
The number of viewers of a programme is taken as the weight of the
vertex. A set of programmes that are mutually exclusive in
respect of time scheduling is called a session. We assume that a
company sets the objective of selecting the popular programmes in
$k$ parallel sessions among different channels so as to make its
commercial advertisement reach the maximum number of viewers, that
is, a company  selects $k$ suitable programme slots simultaneously
for advertisement. The aim of the paper is, therefore, to {help}
the companies to select the programme slots, which are mutually
exclusive with respect to the time schedule of telecasting time,
in such a way that the total number of viewers of the selected
programme in $k$ parallel slots rises to the optimum level. It is
shown that the solution of this problem is obtained by solving the
maximum weight $k$-colouring problem on an interval {graph}. An
algorithm is designed to solve this just-in-time optimization problem using
$O(kMn^2)$ time, where $n$ and $M$ represent the total number of
programmes of all channels and the upper bound of the viewers of
all programmes of all channels  respectively. The problem considered in this
paper is a daily life problem which is modeled  by  $k$-colouring problem on interval graph.

\noindent {\bf Keywords:} \, Graph theory, Modelling, design and analysis of
algorithms, graph colouring, interval graphs.

\noindent 2000 AMS Subject Classifications: Primary codes: 05C15;
05C17; 05C85; 05C90; Secondary codes:  68R10; 68W05

\section{Introduction}
Today television has acquired the central position of all {our
means of entertainment}. Television is not only the most popular
technological device of entertainment but also the best media for
sending information in the simplest way. Various production units
and advertisement agencies are connected with television. In
recent years, most of the world wide mass is being influenced by
advertisement. The daily requirements of man is now being governed
by the attractive advertisements. \ccc{ The production companies
are like to promote their products through different television
channels taking the help of advertising companies (ACs) (the
companies those are interested to advertise for their parties
products). The production companies are spending a large amount of
money for the purpose. }
 The \ccc{ACs} are adopting
attractive ideas to catch the viewers' attention. The main
objective of the ACs are to catch maximum viewers so that the sale
of \ccc{the corresponding} product becomes maximum.

There are hundreds of channels like CNN, HBO, STAR, ZEE, MGM,
NATIONAL GEOGRAPHIC, AXN, etc., running numerous programmes for 24
hours. Among all these programmes there are some which are very
popular. The programmes like `Guiness World Record Prime Time' at
9:00 a.m. shown at AXN, `Charlie Chaplin' at 9:00 a.m. on ZEE
ENGLISH, `Mission Everest' at 10:00 a.m. on National Geographic
channel and many others are found very much popular. Now the
problem arises how and where an AC relays its advertisement to
make it viewed by large mass of population. Sometimes, depending
on the popularity of the programmes, viewers get divided. Like, in
AXN a serial named `Bay Watch' shown at 8:30 p.m. is very popular
and at the same time the programme `Cindrella' on CNN is also very
popular. So,  there arises a problem of selecting the channel. If
the AC is interested in both the slots {it} \ccc{is} not be in
maximum profit as the viewers get divided.

Nowadays, most of the channels run for 24 hours a day. Suppose BBC
has programmes in the respective schedules such as \ccc{during}
7:00-8:00 hours for a movie, 8:00-8:30 hours for news, and so on.
Again CNN runs its programmes \ccc{during} 0:00-3:00 hours for a
movie, 3:00-4:00 hours for a serial, and so on. Similarly, other
channels are engaged with programmes  with definite slots. For
graphical representation we consider each slot or programme as an
interval on a real line. The interval can be represented as closed
interval $[s_i,f_i]$, where $s_i$ and $f_i$ represent respectively
the starting and the finishing time of the programme.

Each programme slot of a particular channel can thus be
represented as an interval on 0 to 24 hours time interval. All the
programme slots of all channels can be represented as a collection
of intervals on the line segment [0,24]. Each interval has an
weight which is equal to the \ccc{average} number of viewers
watching the corresponding programme. {This} set of intervals
{forms} a weighted interval graph $G$. An {\it interval graph} is
a graph whose vertices can be mapped into unique intervals on the
real line such that two vertices in the graph are adjacent if and
only if their corresponding intervals intersect. An interval graph
is called {\it weighted} if its vertices have weights. Now the
maximum number of viewers  is equal to the weight of the
$k$-colourable subgraph of the interval graph $G$. So this problem
can be modelled as an interval graph.

Interval graphs have been extensively studied and used as models
for many real world problems. The interval graph is one of the
most useful discrete mathematical structure for modelling problems
arising in the real world. It has many applications in various
fields like  archaeology, molecular biology, genetics, psychology,
computer scheduling, storage information retrieval, electrical
circuit design, traffic planning, VLSI design, etc
\cite{gol80,rob78}. Interval graphs have been studied from both
the theoretical and algorithmic points of view.

Various algorithmic problems concerning graphs
 in general and the
graph colouring problem have been solved over last few years {
\cite{blo08,bui08,car08,chi10,gal08,
lew09,lu10,luc06,mal08,mal10,men06,plu10,tal08,zuf08}}.

Two graph colouring problems considered in the
literature are:\\
 (i) the problem of finding minimum number of colours to colour all
the vertices of a graph $G$ so that no two adjacent vertices have
the same colour. This is known as minimum colouring problem
or optimal colouring problem,\\
(ii) given $k$ colours, the problem of finding maximum weight
$k$-colourable subgraph of $G$.

In this paper, a set of television programs are considered as a set of intervals on a real
line and it is shown that these intervals form an interval graph. The proposed daily life problem
then modeled by $k$-colouring problem on interval graph.
An algorithm is designed to solve this just-in-time optimization problem using
$O(kMn^2)$ time, where $n$ and $M$ represent the total number of
programmes of all channels and the upper bound of the viewers of
all programmes of all channels  respectively. This a very suitable application of $k$-colouring problem on interval graph.

\subsection{Survey of related works}
 For arbitrary graphs, above problems are NP-complete \cite{kar72}.
 A great deal of research has been
focussed to identify classes of graphs for which these problems
are solvable in polynomial time. Chordal graphs
\cite{bal91,bod93,ho88,nao89,yan87}, interval graphs
\cite{gol80,gup82,gya85,ola91,slu89}, planar graphs \cite{boy87},
 outer planar graphs \cite{dis89}, trees
\cite{raj92}, etc. are such classes.

The maximum weight $k$-colourable subgraph problem in chordal
graphs is polynomially solvable when $k$ is fixed and NP-hard when
$k$ is not fixed \cite{yan87}. The maximum-weight independent set
and maximum-weight $k$-independent set problems are also studied
on many {graph} classes like permutation graphs \cite{sah03,
sah05}, trapezoid graphs \cite{hot01}, circular-arc graphs
\cite{man02}, etc. Efficient algorithms are designed to solve the
maximal independent {set problems} on trapezoid graphs
\cite{hot99} and permutations graphs \cite{pal98}. A minor
variation of minimum colouring problem, called mutual exclusion
scheduling is recently studied for interval graphs \cite{gar06},
permutation graphs \cite{jan03} and comparability graphs
\cite{jan03}.

 A parallel algorithm {has} been presented by Naor {\it et al.} \cite{nao89} to solve optimal colouring problem for chordal
{graphs}. Assuming that all the maximum cliques are part of the
input, this algorithm runs in $O(\log^{2}n)$ time using $O(n^{3})$
processors.

 Olariu \cite{ola91} has solved the optimal colouring problem for interval
graphs in $O(n+m)$ time, where $n$ and $m$ are the number of
vertices and edges respectively, using  greedy heuristic
technique. Pal and Bhattacharjee \cite{pal96} have designed a
parallel algorithm to solve optimal colouring problem on interval
graphs which takes $O(n/P+\log n)$ time {using} $P$ processors.

Recently, just-in-time schedule algorithms have been designed to solve flow-shop problem \cite{sha12}, two-machine flow shop problem \cite{ela13}, single machine flow shop \cite{mos13}, etc.  Kovalyov {\it et al.} \cite{kov07} have discussed about the theory  of fixed interval scheduling.

Recently, Saha, Pal and Pal \cite{sah07} have {solved} 1-colouring
problem for an interval graph whose vertex weights are taken as
interval numbers. The proposed algorithm takes $O(n)$ time, for an
interval graph with $n$ vertices. This algorithm has been applied
to solve the problem that involves selecting different programme
slots (for a single session) telecast on different television
channels in a day so as to reach the maximum number of viewers. In
this problem they have assumed that the number of viewers of the
programme slots are interval numbers.

\subsection{The result obtained}
In this paper, we have designed an algorithm to solve maximum
weight $k$-colouring (MWkC) problem on interval graphs which takes
$O(kMn^2)$ time, where $n$ is the number of vertices and $M$ is
the upper bound on the weights of vertices. A set of programmes
that are mutually exclusive in respect of time scheduling is
called a {{\it session}}. This algorithm is used to select optimal
programme slots which {run} in $k$ parallel sessions such that the
total number of viewers of the selected programmes is maximum. {In
this paper, it is proved that} such programme slots can be
selected using $O(kMn^2)$ time, where $n$ is the total number of
programmes telecasting in all channels during 24 hours and $M$ is
the least upper bound of the viewers among all programmes.

{An outline to solve the proposed problem is given below.}

{The original problem is stated clearly as Problem P1 \ccc{in
Section 2}. It is explained that this problem can be modelled as
an interval graph $G$ and the solution of MWkC problem on $G$ is
the solution of P1. This problem is stated as Problem P2. To solve
the problem P2, a network $N$ is constructed. \ccc{The
conventional} $k$-flow problem on $N$ is stated as Problem P3.
\ccc{This} network flow problem determines the \ccc{minimum}
weight $k$-flow. The maximum weight $k$-flow problem is stated as
P4 \ccc{and this problem is equivalent to the problem P2}.
Unfortunately, no suitable algorithm is available to solve maximum
weight $k$-flow problem on a network. Thus, the maximum weight
$k$-flow problem on $N$ is converted to a minimum weight $k$-flow
problem on the network $N^U$ by applying a suitable
transformation. This transformed problem is defined as Problem
P5.}

{During the conversion, it is proved that \ccc{the problem P1 is
equivalent to P2, P2 is equivalent to P4 and P4 is equivalent to
P5}. Finally, the solution of the problem P1 is obtained from the
solution of the problem P5. }

\section{Modelling of the Problem}
 Interval graphs are useful and significant in the process of
modelling many real life situations, specially involving time
dependencies. In this problem, we represent a programme slot as an
interval. These slots or schedules of the programmes of a channel
are denoted as vertices. If there exist intersection of timings in
between two or more channels; this intersection is regarded as an
edge between the vertices. If finishing time of a programme is the
starting time of another programme then we assume that these
programmes are non-intersecting. It may be noted that any two
programmes in a particular television channel are
non-intersecting. Various programmes have certain number of
viewers which can be determined by statistical survey. The number
of viewers of each programme is considered the weight of the
corresponding {vertex of the} interval graph.

A {\it colouring} of a graph is an assignment of colours to its
vertices so that no two adjacent vertices have the same colour.
The vertices of one colour form a {\it colour class}. Any two
vertices of a colour class are not adjacent. A $k$-colouring of a
graph $G$ uses $k$ colours. The {\it chromatic number} $\chi$ is
defined as the minimum $k$ for which $G$ has a $k$-colouring. A
graph $G$ is $k$-colouring if $\chi\le k$ and is $k$-chromatic if
$\chi=k$. A $k$-colouring of a $k$-chromatic graph is an {\it
optimal colouring}. The weight of a $k$-colourable subgraph (this
subgraph is $k$-chromatic) is the sum of weights of all vertices
of the subgraph. Maximum weight $k$-colouring (MWkC) problem is to
find a $k$-chromatic subgraph whose weight is maximum among all
other $k$-chromatic subgraphs.

If a graph is $k$-chromatic then its vertex set $V$ can be
partitioned into $k$ disjoint sets. Let $V=\{H_1, H_2, \ldots,
H_k\}$ be such a partition, where $H_i \cap H_j=\phi$, for all $i,
j=1, 2, \ldots, k$, $i \not= j$, and for all $u, v\in H_i, i=1, 2,
\ldots, k$, $(u,v)\not\in E$ but if $u\in H_i$ and $v\in H_j$,
{there may be an edge between $u$ and $v$}. Thus, the colour $i$
can be assigned to the set $H_i$. These sets $H_i$, $i=1, 2,
\ldots, k$ are called chromatic partitions of $V$.

We first assume that  AC wishes to run his advertisement from 0 to
24 hours in one session, that is, in a single duration the
advertisement is to be shown in only one channel. This is
equivalent to the maximum weight colouring problem on interval
graph. The problem is also known as maximum weight 1-colouring
problem (MW1C). But, if the company is interested to run {the}
advertisement
 simultaneously in
two parallel sessions it becomes maximum weight 2-colouring
problem (MW2C). {Hence} for $k$ parallel sessions it is termed as
maximum weight $k$-colouring problem (MWkC). Now our problem is to
determine the maximum weight subgraph which can be coloured using
exactly $k$ colours. {The formal definition of the problem is
given below}.

\bigskip
\noindent {\bf Problem P1:} {\sl Suppose a company or any
organization is interested \ccc{to} run \ccc{its} advertisement
simultaneously in $k$ parallel sessions by selecting some
{programme slots}. The restriction is that, any two programmes in
the same session are disjoint. The objective of the problem is to
select {some} {programme slots} such that the sum of the viewers
of the selected programmes is maximum.}

\medskip
\ccc{It is easy to observed that all the programme slots of all
channels in a geographical area can be represented as a circular
arc graph, which is a super class of interval graph. Here we
assumed that at some point of time in a day (say, at 0:00 hours at
midnight), all programmes that broadcasted earlier must terminate
at or before 0:00 hours and all programmers in all channels would
start broadcasting exactly at or after 0:00 hours and no programme
would start earlier to 0:00 hours and continue after 0:00 hours. }

 So the
{\bf Problem P1} can be solved by solving the following {Problem
P2} on the constructed interval graph.

\bigskip
\noindent {\bf Problem P2:} {\sl Find a subgraph $H(G)$ of an
interval graph $G$ which can be coloured using exactly $k$ colours
such that $\sum_{u\in H(G)} w(u)$ is maximum among all other such
{subgraphs}, where $w(u)$ is the weight of the vertex $u$.}

\medskip
This problem is referred to as the maximum weight $k$-coluring
problem. It is easy to observe that the Problems P1 and P2 are
equivalent.

 In the following sections,  the solution {procedure}
of the problem MWkC {is discussed}.

\section{Cliques of Interval Graph}
A {\it clique} of a graph is a set of vertices such that every two
vertices in the set are joined by an edge. An {\it independent set}
of a graph is the set of vertices such that any two vertices in
the set are not connected by an edge. In a colouring, each colour
class is an independent set, so $G$ is $k$-colourable if and only
if $V$ is the union of $k$ independent sets. Thus,
`$k$-colourable' and `$k$-partite' (a graph is {\it $k$-partite}
if its vertices can be expressed as the union of $k$ independent
sets) have the same meaning. The usage of the two terms are
slightly different. The `$k$-partite' is a structural hypothesis,
while `$k$-colourable' is the result of an optimization problem.

 Let $G=(V,E)$, $V=\{1, 2,\ldots, n\}$ be an interval graph and $\cal{C}$ $= \{C_1, C_2, \ldots,
C_r\}$, for some $r$, be all maximal cliques.  The maximal cliques
of an interval graph satisfy the following result.

\begin{lemma}
\label{l10}
The maximal cliques of an interval graph $G$ can be linearly ordered such
that, for every vertex $u\in G$, the maximal cliques containing $u$ occur
consecutively \cite{gol80}.
\end{lemma}

In an interval graph the {\it leading point} of a clique is the
leftmost left endpoint of the interval at which all the other
intervals in that clique intersect. {It is assumed} that the
cliques are in order of their increasing leading point. It has
been shown in \cite{gol80} that for perfect graph the clique
number $\alpha$ (the size of a maximum clique) is equal to the
chromatic number $\chi$. Therefore, the vertex set $V$ can be
partitioned into \ccc{$\alpha$} disjoint sets such that each of
them is an independent set, i.e., $\ccc{V=\cup_{i=1}^{\alpha}}
V_i$, where each $V_i$ is an independent set and $V_i\cap
V_j=\phi, i\not=j$. Thus, if $k\ge \alpha$ then $|V|$ is the
maximum number of vertices in $k$-independent set. All the
vertices of an independent set can be coloured by a single colour.
Since interval graphs are perfect, therefore MWkC may be found by
locating a maximum weight subgraph among all subgraphs \ccc{by
$k$-colouring}. We propose to find such a subgraph for the
interval graph.

\begin{figure}
\special{em:linewidth 0.4pt}
\unitlength 1.00mm
\linethickness{0.4pt}
\begin{center}
\begin{picture}(105.33,31.34)
\emline{5.33}{3.34}{1}{5.33}{6.00}{2}
\emline{10.33}{3.34}{3}{10.33}{6.00}{4}
\emline{15.33}{6.00}{5}{15.33}{3.34}{6}
\emline{20.33}{3.34}{7}{20.33}{6.00}{8}
\emline{25.33}{3.34}{9}{25.33}{6.00}{10}
\emline{30.33}{6.00}{11}{30.33}{3.34}{12}
\emline{35.33}{3.34}{13}{35.33}{6.00}{14}
\emline{40.33}{6.00}{15}{40.33}{3.34}{16}
\emline{45.33}{3.34}{17}{45.33}{6.00}{18}
\emline{50.33}{6.00}{19}{50.33}{3.34}{20}
\emline{55.33}{3.34}{21}{55.33}{6.00}{22}
\emline{60.33}{6.00}{23}{60.33}{3.34}{24}
\emline{65.33}{3.34}{25}{65.33}{6.00}{26}
\emline{70.33}{6.00}{27}{70.33}{3.34}{28}
\emline{75.33}{3.34}{29}{75.33}{6.00}{30}
\emline{80.33}{6.00}{31}{80.33}{3.34}{32}
\emline{85.33}{3.34}{33}{85.33}{6.00}{34}
\emline{90.33}{6.00}{35}{90.33}{3.34}{36}
\emline{95.33}{3.34}{37}{95.33}{6.00}{38}
\emline{100.33}{6.00}{39}{100.33}{3.34}{40}
\emline{5.33}{8.34}{41}{30.33}{8.34}{42}
\emline{35.33}{8.34}{43}{40.33}{8.34}{44}
\emline{60.33}{8.34}{45}{75.33}{8.34}{46}
\emline{80.33}{8.34}{47}{100.33}{8.34}{48}
\put(17.33,11.00){\makebox(0,0)[cc]{$I_2(3)$}}
\put(37.66,11.00){\makebox(0,0)[cc]{$I_3(6)$}}
\put(68.33,11.00){\makebox(0,0)[cc]{$I_6(4)$}}
\put(90.33,11.00){\makebox(0,0)[cc]{$I_{10}(3)$}}
\emline{10.33}{15.00}{49}{15.33}{15.00}{50}
\emline{20.33}{15.00}{51}{55.33}{15.00}{52}
\emline{65.33}{15.00}{53}{95.33}{15.00}{54}
\put(13.00,17.34){\makebox(0,0)[cc]{$I_1(5)$}}
\put(80.33,17.34){\makebox(0,0)[cc]{$I_9(5)$}}
\put(37.66,17.67){\makebox(0,0)[cc]{$I_5(8)$}}
\emline{25.33}{21.34}{55}{45.33}{21.34}{56}
\emline{50.33}{21.34}{57}{80.33}{21.34}{58}
\put(36.00,24.00){\makebox(0,0)[cc]{$I_4(2)$}}
\put(65.66,24.00){\makebox(0,0)[cc]{$I_7(1)$}}
\put(80.33,31.34){\makebox(0,0)[cc]{$I_8(2)$}}
\put(5.33,0.67){\makebox(0,0)[cc]{1}}
\put(25.33,0.67){\makebox(0,0)[cc]{5}}
\put(50.33,0.67){\makebox(0,0)[cc]{10}}
\put(75.33,0.67){\makebox(0,0)[cc]{15}}
\put(100.33,0.67){\makebox(0,0)[cc]{20}}
\emline{85.33}{21.34}{59}{50.33}{21.34}{60}
\emline{70.33}{27.67}{61}{90.33}{27.67}{62}
\emline{0.33}{4.67}{63}{105.33}{4.67}{64}
\put(50.33,-5.00){\makebox(0,0)[cc]{(a) A set of intervals. The numbers
within parentheses represent weights.}}
\end{picture}

\vspace{1cm}
\special{em:linewidth 0.4pt}
\unitlength 1mm
\linethickness{0.4pt}
\begin{picture}(82.01,51.33)
\put(4.67,21.67){\circle{4.67}} \put(24.67,21.67){\circle{4.67}}
\put(14.67,36.67){\circle{4.67}} \put(14.67,8.67){\circle{4.67}}
\put(48.67,21.67){\circle{4.67}}
\emline{46.33}{21.67}{1}{27.00}{21.67}{2}
\emline{22.33}{21.67}{3}{7.00}{21.67}{4}
\emline{4.67}{19.33}{5}{12.33}{8.33}{6}
\emline{17.00}{8.33}{7}{24.67}{19.33}{8}
\emline{12.33}{36.67}{9}{4.33}{24.00}{10}
\emline{17.00}{36.33}{11}{24.67}{24.00}{12}
\put(50.33,8.67){\circle{4.67}} \put(69.00,12.00){\circle{4.67}}
\put(69.00,30.67){\circle{4.67}} \put(79.67,45.67){\circle{4.67}}
\emline{77.67}{45.67}{13}{77.33}{46.00}{14}
\emline{51.00}{21.33}{15}{67.00}{30.67}{16}
\emline{79.33}{43.33}{17}{70.67}{32.67}{18}
\emline{69.00}{14.33}{19}{69.00}{28.33}{20}
\emline{52.67}{8.67}{21}{66.67}{12.00}{22}
\emline{48.67}{19.00}{23}{50.33}{10.67}{24}
\emline{51.67}{10.67}{25}{67.33}{29.00}{26}
\emline{51.00}{21.33}{27}{67.33}{13.33}{28}
\emline{79.33}{43.00}{29}{69.00}{14.33}{30}
\put(23.67,46.00){\circle{4.67}}
\emline{14.67}{39.00}{31}{23.67}{43.67}{32}
\put(38.00,2.00){\makebox(0,0)[cc] {(b) The interval graph
corresponding to the above intervals.}}
\emline{48.67}{24.00}{33}{79.33}{43.33}{34}
\put(23.67,46.00){\makebox(0,0)[cc]{1}}
\put(24.00,51.33){\makebox(0,0)[cc]{5}}
\put(14.33,37.00){\makebox(0,0)[cc]{2}}
\put(10.00,39.00){\makebox(0,0)[cc]{3}}
\put(4.67,21.67){\makebox(0,0)[cc]{4}}
\put(1.33,26.33){\makebox(0,0)[cc]{2}}
\put(14.67,8.67){\makebox(0,0)[cc]{3}}
\put(9.67,6.00){\makebox(0,0)[cc]{6}}
\put(24.67,21.67){\makebox(0,0)[cc]{5}}
\put(27.00,26.33){\makebox(0,0)[cc]{8}}
\put(48.67,21.67){\makebox(0,0)[cc]{7}}
\put(46.67,25.67){\makebox(0,0)[cc]{1}}
\put(79.67,45.67){\makebox(0,0)[cc]{10}}
\put(78.00,50.33){\makebox(0,0)[cc]{3}}
\put(69.00,30.67){\makebox(0,0)[cc]{9}}
\put(69.00,34.67){\makebox(0,0)[cc]{5}}
\put(50.33,8.67){\makebox(0,0)[cc]{6}}
\put(45.67,11.00){\makebox(0,0)[cc]{4}}
\put(69.00,12.00){\makebox(0,0)[cc]{8}}
\put(72.67,15.00){\makebox(0,0)[cc]{2}}
\end{picture}
\end{center}
\caption{\label{f10} A set of intervals and its corresponding
interval graphs.}
\end{figure}

For the graph of Figure \ref{f10}(b), the maximal cliques are $C_1=\{1,2\}$,
$C_2=\{2,4,5\}$, $C_3=\{3,4,5\}$, $C_4=\{5,7\}$, $C_5=\{6,7,8,9\}$,
$C_6=\{7,8,9,10\}$.

The MWkC problem is solved by converting the problem to an
equivalent problem on a network (Directed Acyclic Graph, in short
DAG). Then solving the problem on DAG, the solution of Problem P2
{ is obtained} and hence the solution of Problem P1. In the
following section, DAG { is defined } and  \ccc{a method is
described} to construct a DAG for an interval graph.

\section{The Network Flow Problem}
A {\it network} $N$ is a finite set of nodes and a subset of the
ordered pairs $(u,v)$, $u\not=v$, called the {\it arcs}. The
network $N$ has a special return arc $(t,s)$, where node $s$ is
called the {\it source} in $N$ and node $t$ is called the {\it
sink} in $N$. The set of all arcs of $N$, except $(t,s)$ is
denoted by $E_N$. Further, a positive real-valued capacity
$c(u,v)>0$ and a non-negative weight $w_N(u,v)$ are associated
with each edge $(u,v)$. For simplicity, {the} capacity for each
{non-existing} edge is assumed to be zero, i.e., $c(u,v)=0$ if
$(u,v)\not\in E$. A flow on $G$ is a real valued function $f$ on
$E_N$ if it satisfies the following three conditions:
\begin{eqnarray*}
&&f(u,v)=-f(v,u), \;\; \mbox{for all}\;\; (u,v)\in E_N,\\
&& f(u,v)\le c(u,v), \;\; \mbox{for all}\;\; (u,v)\in E_N,\\
&&\sum_v f(u,v)=0,  \;\; \mbox{for all}\;\; v\in V-\{s,t\}.
\end{eqnarray*}

For each arc $(u,v)$ of $E_N$, $f(u,v)$ represents the amount of flow in
the arc $(u,v)$.

For a network $N$ the minimum weight $k$-flow problem is defined for  a
network
$N$ as follows:

\bigskip
\noindent {\bf Problem P3:} {\sl The minimum weight $k$-flow
problem is to obtain $k$ edge disjoint paths $P_1, P_2, \ldots,
P_k$ from the set of all possible paths from $s$ to $t$ in $N$ to
$$minimize\,\, \sum_{(u,v)\in J_k} w_N(u,v)f(u,v)$$
where\\
(i) $f(u,v)=0$ or 1, for $(u,v)\in E_N$,\\
(ii) $J_k=\cup_{i=1}^kE_N^i$,\\
(iii) $E_N^i$ is the set of arcs associated with the path $P_i$}.

For each arc $(u,v)$ of $E_N$, $f(u,v)$ represents the amount of
flow in the arc $(u,v)$, and also it represents the net amount of
flow from $v$ to $u$ in the rest of the network ``$N-(u,v)$''.

\section{Construction of the Network}
Let us consider a network $N$ with nodes $C_0\; (=s), C_1, C_2, \ldots,
C_r\; (=t)$ and arcs $(C_{i-1}, C_i)$, $i=1, 2, \ldots, r$, where
$C_0$ is empty. Let each of these arcs, called a {\it $c$-arc}, be
given a weight 0 and a capacity $k$.

By Lemma \ref{l10} for each $u\in V$ there exist consecutive cliques $C_p,
C_{p+1},
\ldots, C_q$, such that $u\in C_p, u\in C_{p+1},\ldots, u\in
C_{q}$, $p\le q$ but, $u\not\in C_{p-1}, u\not\in C_{q+1}$.
Here we note that for $u\in V$ we have $u\in C_p$, $u\in
C_q$ and $p\le q$ but $u\not\in C_i$ for any
$i<p$ and $u\not\in C_j$ for any $j>q$. We add an arc $(C_{p-1}, C_q)$ to the
network $N$ and assign the weight $w(u)$ and capacity 1 to this
arc. For each vertex $u\in V$, we get such an arc.
Let these arcs be called the {\it $i$-arcs}. Let $V_N$ and
$E_N$ be the set of nodes and the set of arcs of the network $N$
respectively. Obviously, the network $N$ is acyclic. Here the
capacity of each $i$-arc is 1 and that of each $c$-arc is $k$. So,
this network is referred also as integral flow network. We may
replace each $c$-arc by $k$ parallel arcs and assigning capacity 1
and weight 0 to each of them.

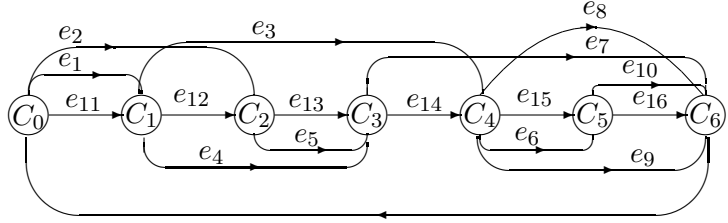
\begin{figure}
\special{em:linewidth 0.4pt}
\unitlength 1.00mm
\linethickness{0.4pt}
\begin{center}
\begin{picture}(97.34,35.34)
\put(4.67,14.67){\circle{5.33}}
\put(19.67,14.67){\circle{5.33}}
\put(34.67,14.67){\circle{5.33}}
\put(49.67,14.67){\circle{5.33}}
\put(64.67,14.67){\circle{5.33}}
\put(79.67,14.67){\circle{5.33}}
\put(94.67,14.67){\circle{5.33}}
\put(49.67,14.67){\makebox(0,0)[cc]{$C_3$}}
\put(64.67,14.67){\makebox(0,0)[cc]{$C_4$}}
\put(79.67,14.67){\makebox(0,0)[cc]{$C_5$}}
\put(94.67,14.67){\makebox(0,0)[cc]{$C_6$}}
\put(4.67,14.34){\makebox(0,0)[cc]{$C_0$}}
\put(19.67,14.67){\makebox(0,0)[cc]{$C_1$}}
\put(34.67,14.67){\makebox(0,0)[cc]{$C_2$}}
\put(32.00,14.67){\vector(1,0){0.2}}
\emline{22.34}{14.67}{1}{32.00}{14.67}{2}
\put(47.00,14.67){\vector(1,0){0.2}}
\emline{37.34}{14.67}{3}{47.00}{14.67}{4}
\put(62.00,14.67){\vector(1,0){0.2}}
\emline{52.34}{14.67}{5}{62.00}{14.67}{6}
\put(77.00,14.67){\vector(1,0){0.2}}
\emline{67.34}{14.67}{7}{77.00}{14.67}{8}
\put(92.00,14.67){\vector(1,0){0.2}}
\emline{82.34}{14.67}{9}{92.00}{14.67}{10}
\put(17.00,14.67){\vector(1,0){0.2}}
\emline{7.34}{14.67}{11}{17.00}{14.67}{12}
\put(12.17,17.50){\oval(15.00,5.00)[t]}
\put(34.84,11.01){\oval(29.67,6.67)[b]}
\put(42.17,12.17){\oval(15.00,4.33)[b]}
\put(72.17,18.84){\oval(45.00,7.00)[t]}
\put(72.17,12.00){\oval(15.00,4.00)[b]}
\put(79.50,11.83){\oval(30.33,9.00)[b]}
\put(87.17,17.34){\oval(15.00,2.67)[t]}
\emline{64.67}{17.34}{13}{66.23}{19.15}{14}
\emline{66.23}{19.15}{15}{67.80}{20.76}{16}
\emline{67.80}{20.76}{17}{69.36}{22.16}{18}
\emline{69.36}{22.16}{19}{70.93}{23.37}{20}
\emline{70.93}{23.37}{21}{72.50}{24.36}{22}
\emline{72.50}{24.36}{23}{74.07}{25.16}{24}
\emline{74.07}{25.16}{25}{75.64}{25.75}{26}
\emline{75.64}{25.75}{27}{77.21}{26.14}{28}
\emline{77.21}{26.14}{29}{78.79}{26.32}{30}
\emline{78.79}{26.32}{31}{80.37}{26.30}{32}
\emline{80.37}{26.30}{33}{81.95}{26.08}{34}
\emline{81.95}{26.08}{35}{83.53}{25.65}{36}
\emline{83.53}{25.65}{37}{85.12}{25.02}{38}
\emline{85.12}{25.02}{39}{86.70}{24.18}{40}
\emline{86.70}{24.18}{41}{88.29}{23.14}{42}
\emline{88.29}{23.14}{43}{89.88}{21.90}{44}
\emline{89.88}{21.90}{45}{91.47}{20.45}{46}
\emline{91.47}{20.45}{47}{94.34}{17.34}{48}
\put(42.17,17.50){\oval(45.67,13.67)[t]}
\put(19.67,17.34){\oval(30.00,12.67)[t]}
\put(16.00,23.67){\vector(1,0){0.2}}
\emline{14.67}{23.67}{49}{16.00}{23.67}{50}
\put(14.34,20.00){\vector(1,0){0.2}}
\emline{12.00}{20.00}{51}{14.34}{20.00}{52}
\put(46.67,24.34){\vector(1,0){0.2}}
\emline{43.34}{24.34}{53}{46.67}{24.34}{54}
\put(44.34,10.00){\vector(1,0){0.2}}
\emline{41.67}{10.00}{55}{44.34}{10.00}{56}
\put(35.34,7.67){\vector(1,0){0.2}}
\emline{31.67}{7.67}{57}{35.34}{7.67}{58}
\put(73.34,10.00){\vector(1,0){0.2}}
\emline{71.00}{10.00}{59}{73.34}{10.00}{60}
\put(82.34,7.34){\vector(1,0){0.2}}
\emline{79.00}{7.34}{61}{82.34}{7.34}{62}
\put(77.34,22.34){\vector(1,0){0.2}}
\emline{74.00}{22.34}{63}{77.34}{22.34}{64}
\put(88.67,18.67){\vector(1,0){0.2}}
\emline{86.00}{18.67}{65}{88.67}{18.67}{66}
\put(80.00,26.33){\vector(1,0){0.2}}
\emline{78.00}{26.33}{67}{80.00}{26.33}{68}
\emline{20.00}{9.67}{69}{20.00}{11.67}{70}
\emline{49.67}{18.34}{71}{49.67}{17.34}{72}
\put(10.34,21.67){\makebox(0,0)[cc]{$e_1$}}
\put(10.34,25.34){\makebox(0,0)[cc]{$e_2$}}
\put(36.00,26.00){\makebox(0,0)[cc]{$e_3$}}
\put(29.00,9.34){\makebox(0,0)[cc]{$e_4$}}
\put(41.67,11.67){\makebox(0,0)[cc]{$e_5$}}
\put(71.00,11.84){\makebox(0,0)[cc]{$e_6$}}
\put(86.34,9.00){\makebox(0,0)[cc]{$e_9$}}
\put(80.34,24.00){\makebox(0,0)[cc]{$e_7$}}
\put(80.00,28.67){\makebox(0,0)[cc]{$e_8$}}
\put(86.00,20.84){\makebox(0,0)[cc]{$e_{10}$}}
\put(12.00,16.34){\makebox(0,0)[cc]{$e_{11}$}}
\put(26.34,17.00){\makebox(0,0)[cc]{$e_{12}$}}
\put(41.67,16.34){\makebox(0,0)[cc]{$e_{13}$}}
\put(57.34,16.34){\makebox(0,0)[cc]{$e_{14}$}}
\put(72.00,17.00){\makebox(0,0)[cc]{$e_{15}$}}
\put(87.34,16.67){\makebox(0,0)[cc]{$e_{16}$}}
\emline{49.67}{19.00}{73}{49.67}{17.34}{74}
\emline{94.67}{18.67}{75}{94.67}{17.34}{76}
\put(49.67,11.67){\oval(90.67,20.67)[b]}
\put(51.00,1.33){\vector(-1,0){0.2}}
\emline{53.33}{1.33}{77}{51.00}{1.33}{78}
\end{picture}
\end{center}
\caption{\label{f20} The network $N$ for the graph of Figure \ref{f10}(b).
The (weight,capacity) of each arc are: $e_1: (5,1), e_2: (3,1), e_3: (8,1),
e_4: (2,1), e_5: (6,1), e_6: (4,1), e_7: (1,1), e_8: (2,1), e_9: (5,1),
e_{10}: (3,1), e_{11}: (0,k), e_{12}: (0,k), e_{13}: (0,k), e_{14}: (0,k),
e_{15}: (0,k), e_{16}: (0,k)$.}
\end{figure}

The network $N$ for the graph of Figure \ref{f10}(b) is shown in
Figure \ref{f20}. The network has seven vertices $C_0, C_1, C_2, \ldots,
C_6$ and sixteen
edges $e_1, e_2, \ldots, e_{16}$ of which $e_1, e_2, \ldots, e_{10}$ are
$i$-arcs and $e_{11}, e_{12}, \ldots, e_{16}$ are $c$-arcs. The (vertex,
weight, capacity) of each $i$-arc of the network
of Figure \ref{f20} is shown below:\\
$e_1: (1, 5,1), e_2: (2, 3,1), e_3: (5, 8,1),
e_4: (4, 2,1), e_5: (3, 6,1), e_6: (6, 4,1), e_7: (7, 1,1), e_8: (8, 2,1),
e_9: (9, 5,1), e_{10}: (10, 3,1)$.

Let us consider the maximum $k$-flow problem {\bf P4} on the network $N$.

\bigskip
\noindent {\bf Problem P4:} {\sl The maximum weight $k$-flow
problem is to find the set of $k$ disjoint paths $P_1, P_2,
\ldots, P_k$ from the set of all possible paths from $s$ to $t$ in
$N$ to
$$maximize \,\, \sum_{(u,v)\in J_k} {w_N(u,v)}f(u,v)$$
where\\
 (i) $J_k=\cup_{i=1}^k E_N^i$,\\
(ii) $E_N^i$ is the set of arcs associated with the path $P_i$,\\
(iii) the value of $f(u,v)$ is either 0 or 1 for all {$(u,v)\in
E_N$}. }

\section{Properties of the Network $N$}
The following result for a triangulated graph, given by Fulkerson
and Gross \cite{ful65}, is also valid for an interval graph,
because an interval graph satisfies the properties of triangulated
graphs.

\begin{lemma}
\label{l20}  A triangulated graph and so an interval graph with $n$
vertices has at most $n$ maximal cliques. The number of maximal
cliques is $n$ if and only if the graph has no edges \cite{ful65}.
\end{lemma}

The nodes of the network $N$ are $C_0, \ldots, C_r$, i.e., total
number of nodes is $r+1$. The arcs are of two types $c$-arc and
$i$-arc. Each vertex is associated with an $i$-arc and each
$c$-arc is drawn to join two consecutive nodes of the network.
Thus the total number of $i$-arc is $n$ and that of $c$-arc is
$r$. Hence, we can conclude the following result.

\begin{lemma}
\label{l30}  The  total  number  of  nodes  and  arcs  of  $N$ are
respectively  $r+1$  and  $n+r+1$  which are of $O(n)$, where
$r<n$.
\end{lemma}

The following lemma establishes the relationship between the arcs
of the network $N$ and the vertices of the interval graph $G$.

\begin{lemma}
\label{l40} If {$(C_i, C_j)$} and {$(C_j, C_l)$} be two
consecutive $i$-arcs of the set $E_N$ then the corresponding
vertices are non-adjacent in $G$.
\end{lemma}

\noindent {\bf Proof.} Let $x$ and $y$ be the vertices
corresponding to the $i$-arcs {$(C_i,C_j)$} and {$(C_j,C_l)$}
respectively. From the definition of $N$, it is clear that $x$
belongs to the cliques $C_{i+1}, C_{i+2}, \ldots, C_j$, but not in
$C_{j+1}$ and $C_i$ and similarly, $y$ {belongs} to the cliques
$C_{j+1}, C_{j+2}, \ldots, C_l$, but not in $C_j$ and $C_{l+1}$.
That is, there is no common clique for $x$ and $y$. Hence,
$(x,y)\not\in E.$ \hfill$\Box$

The following lemma gives the guarantee about the existence of a
$k$-flow in the network $N$.

\begin{lemma}
\label{l50} The network $N$ has a $k$-flow.
\end{lemma}

\noindent {\bf Proof.} Let $B_1={\{C_0, C_1, \ldots, C_i\}}$ and
$B_2={\{C_{j}, C_{j+1}, \ldots, C_r\}}$ be two sets of vertices,
for a given $i$, $1\le i \le r$, where $j=i+1$.

\noindent {\it Case 1:} Let there be no $i$-arcs between the nodes
of $B_1$ and $B_2$.\\
In this case every flow passes through the nodes ${C_i}\in B_1$
and ${C_j}\in B_2$ and the only connected arc is a $c$-arc with
capacity $k$. Then there are at most $k$ flows from $s$ to $t$
through {$C_i, C_j$}, because the flow of each $i$-arc of $N$ is
either 0 or 1.

\noindent {\it Case 2:} Let there be at least one $i$-arc between
the nodes of $B_1$ and $B_2$.

In this case, all flows do not necessarily pass through {$C_i$}
and {$C_j$}. That is, there is at least one flow from a vertex of
$B_1$ to a vertex of $B_2$ except {$C_i$} and {$C_j$} and $k$
flows from {$C_i$} to {$C_j$}. Therefore, there is at least $k+1$
flows from $s$ to $t$. \hfill $\Box$

The following theorem proves that the problems P2 and P4 are
equivalent.

\begin{theorem}
\label{t10} The problem P2 for $G$ and problem P4 for $N$ are
equivalent.
\end{theorem}

\noindent {\bf Proof.} The construction of the network $N$ from the
weighted graph $G$ shows that there is one to one correspondence
between the set of vertices of $G$ and the set of $i$-arcs of $N$.
Also each $i$-arc corresponds to a vertex of $G$ and the vertices
corresponding to the $i$-arcs at the end of any $c$-arc are not
directly connected. Hence each path from $s$ to $t$ in $N$
corresponds to an independent set of $G$ and conversely, each
maximal independent set of $G$ corresponds to a path from $s$ to $t$
in $N$. Further, we note that the weight of any $c$-arc is 0 and the
weight of any $i$-arc is same as the weight of the corresponding
vertex of $G$. Thus, the total weight of any $k$ colour classes of
$G$ and the total weight of the arcs of $k$ paths of $N$ are same.
We take $f(x,y)=1$ for each arc associated with these $k$ paths and
$f(x,y)=0$ otherwise. Let the sets $H_1, H_2, \ldots, H_k$ of
vertices of Problem P2 correspond to the set of paths $P_1, P_2,
\ldots, P_k$ of Problem P4 respectively. Let $I_k=\cup_{i=1}^k H_i$.

Therefore,
\begin{eqnarray*}
\sum_{v\in I_k} w(v) &=& \sum_{i=1}^k \sum_{v\in H_i} w(v)\\
&=& \sum_{i=1}^k \sum_{(x,y)\in E_N^i} w_N(x,y)\\
&&\;\; ({\rm as}\;\; w_N(x,y)=w(v)\;\; {\rm for}\;\; {{\it i}-{\rm
arc}}\;\;
and\\
&&  w_N(x,y)=0\;\; {\rm for}\;\; {{\it c}-{\rm arc}})\\
&=& \sum_{i=1}^k \sum_{(x,y)\in E_N^i} w_N(x,y) f(x,y)\\
&& \;\; (\mbox{since for all}\;\; i, f(x,y)=1,\;\; \mbox{for}\;\; \\
&& (x,y)\in E_N^i,\;\; {\rm and}\;\; f(x,y)=0,\;\; {\rm
       otherwise})\\
&=&\sum_{(x,y)\in J_k} w_N(x,y) f(x,y).
\end{eqnarray*}

Hence problems P2 and P4 are equivalent. \hfill $\Box$






Unfortunately, no algorithm is available to solve {the} maximum
weight $k$-flow problem. But, the minimum weight $k$-flow problem
for $N$ can be solved using the algorithm of Edmonds and Karp
\cite{edm72}. Thus we have to modify $N$ by negating the weight of
each arc and then finding a minimum weight $k$-flow.
Unfortunately, the algorithm of Edmonds and Karp also requires
that all arc weights are non-negative.

In order to convert a maximum weight flow problem to a minimum
weight flow problem with positive arc weight a transformation {is
required}.

To transform the problem, {the} array $\pi$ {is defined} as follows:\\
$\pi(v_i)=$ largest weight of the path from $v_i$ to $t$ in $N$,
$v_i\in V_N, i=0, 1, 2, \ldots, r.$ The array $\pi$ is computed by
the Algorithm $\Pi$.

\bigskip
\noindent {\sc Algorithm $\Pi$}\\
{\it Input:} The network $N$.\\
{\it Output:} The array $\pi(v_i),\;\; v_i\in V_N.$\\
{\it Initialization:} $\pi(v_i)=0, \;\; i=1, 2, \ldots, r$.\\
{\bf for} $i=r-1$ {\bf to} 0 {\bf step} $-1$ {\bf do}\\
\tab {\bf for} each arc $(v_i,v_j)\in E_N$ {\bf do}\\
\tab\tab $\pi(v_i)=\max\{\pi(v_i), w_N(v_i,v_j)+\pi(v_j)\}$;\\
\tab {\bf endfor;}\\
{\bf endfor;}\\
{\bf end $\Pi$}

\bigskip
The time complexity to calculate the array $\pi$ is presented in the
following lemma.

\begin{lemma}
\label{l8} The array $\pi$ can be computed correctly in $O(n)$ time.
\end{lemma}

\noindent {\bf Proof.} Let $m_i$ be the total number of arcs
adjacent to $v_i$. From Algorithm $\Pi$ it follows that the time
complexity of this algorithm is $\sum_{i=1}^{r-1} m_i =$ total
number of arcs of $N$, which is equal to $O(n)$ (from Lemma
\ref{l30}). {As} the network $N$ is acyclic therefore, the
correctness follows from the algorithm directly. Hence, the lemma
follows. \hfill $\Box$

The array $\pi$ for the network $N$ of Figure \ref{f20} is
$\pi(C_0)= 20, \pi(C_1)= 15, \pi(C_2)= 13, \pi(C_3)= 7, \pi(C_4)=
7, \pi(C_5)= 3, \pi(C_6)= 0.$

\section{Construction of a Minimum Weight Flow Network}
Now, we convert the problem {\bf P4} to a minimum weight $k$-flow
problem {\bf P5} as follows:\\
A network $N^U$ is constructed from $N$ using the same set of arcs
$(E_N)$, the same set of nodes $(V_N)$ and identical capacities,
but different weights. The weight on the arc $(v_i,v_j),\;\; i<j$,
is
$$w_{N^U}(v_i,v_j)=\pi(v_i)-\pi(v_j)-w_N(v_i,v_j),$$
for all $(v_i,v_j)\in E_N.$

The problem {\bf P5} is defined as follows:

\bigskip
\noindent
{\bf Problem P5:} {\sl For the network $N^U$ find the
set of $k$ disjoint paths $P_1, P_2, \ldots, P_k$ from the set of
all possible paths from $s$ to $t$ in $N^U$ to
$$minimize \,\, \sum_{(u,v)\in J_k}\; {w_{N^U}}(u,v)f(u,v)$$ where\\
(i) $J_k=\cup_{i=1}^k E_N^i$,\\
(ii) $E_N^i$ is the set of arcs associated with the path $P_i$,\\
(iii) the value of $f(u,v)$ is either 0 or 1 for all $(u,v)\in
{E_N}$.}

\medskip

The Table \ref{tab10} shows the weights of each arc of the
networks $N$ and $N^U$.

\begin{table}
\begin{center}
\begin{tabular}{|l|c|c|c|c|c|c|c|c|c|c|c|c|c|c|c|c|}
\hline
  arcs             & $e_1$ & $e_2$ & $e_3$ & $e_4$ & $e_5$ & $e_6$ & $e_7$ & $e_8$ & $e_9$ & $e_{10}$ & $e_{11}$ & $e_{12}$ & $e_{13}$ & $e_{14}$ & $e_{15}$ & $e_{16}$ \\
\hline
  weights in $N$   & 5 & 3 & 8 & 2 & 6 & 1 & 2 & 5 & 3 & 0 & 0 & 0 & 0 & 0 & 0 & 0\\
\hline
  weights in $N^U$ & 0 & 4 & 0 & 5 & 0 & 0 & 6 & 5 & 2 & 0 & 5 & 2 & 6 & 0 &
4 & 3 \\ \hline
\end{tabular}
\end{center}
\caption{\label{tab10} Weights of arcs of the networks $N$ and
$N^U$.}
\end{table}

The following lemma establishes that the weight of each arc of the
network $N^U$ are non-negative.

\begin{lemma}
\label{l90} The weights $w_{N^U}(v_i,v_j), i<j$ of all arcs of the
network $N^U$ are non-negative.
\end{lemma}

\noindent {\bf Proof.} If possible let the weight $w_{N^U}(v_i,v_j)$
of the arc $(v_i,v_j)$ be negative. From the definition of $N$ it
follows that, in the left to right ordering of the cliques, the
clique corresponding to $v_i$ lies to the left of the clique
corresponding to $v_j, i<j$. Since, $w_{N^U}(v_i,v_j)<0$, therefore
$$\pi(v_i)-\pi(v_j)-w_N(v_i,v_j)<0, \;\; {\rm or}\;\;
\pi(v_i)<\pi(v_j)+w_N(v_i,v_j).$$

But, $\pi(v_i)$ is the largest weight of a path from $v_i$ to $t$
in $N$ and similar is the interpretation for $\pi(v_j)$, so
$$\pi(v_i)\ge\pi(v_j)+w_N(v_i,v_j), i<j.$$
This contradicts the statement made earlier. Hence,
$w_{N^U}(v_i,v_j)\ge 0,$ for all $(v_i, v_j)\in E_N$. \hfill
$\Box$

From the definition of $\pi$ it is clear that $\pi(s)$ is the
largest weight of a path from $s$ to $t$ and $\pi(t)=0$. Thus, for
any arc $(v_i, v_j)\in E_N$, the upper bound of $w_{N^U}(v_i,
v_j)$ is $\pi(s)$, which is proved in the following lemma.

\begin{lemma}
\label{l100} The maximum value of weights of all arcs of the
network $N^U$ is $\pi(s)$,\\ i.e., $\max_{i,j}
w_{N^U}(v_i,v_j)=\pi(s).$
\end{lemma}

\noindent {\bf Proof.} For {$i\le j$},
\begin{eqnarray*}
  \max_{(v_i,v_j)\in E_N} w_{N^U}(v_i,v_j) & =& \max_{(v_i,v_j)\in E_N}
  \{\pi(v_i)-[\pi(v_j)+w_N(v_i,v_j)]\}\\
  &<&\max_{(v_i,v_j)\in E_N} \pi(v_i) = \pi(s),
\end{eqnarray*}
since $\pi(v_j)+w_N(v_i,v_j)\ge 0.$ Thus, the upper bound of the
weights of any arc of the set $E_N$ is the largest weight of a
path from $s$ to $t$ in $N$. Hence, the lemma. \hfill $\Box$

It can be shown that the maximum flow of $N$ is equivalent to
minimum flow of $N^U$.

\begin{lemma}
\label{l110} For the same set of arcs the maximum weight flow from
$s$ to $t$ in $N$ is equal to the minimum weight flow in $N^U$.
\end{lemma}

\noindent {\bf Proof.} Without loss of generality, we assume that the
sequence of arcs of a flow from $s$ to $t$ is $E'_N=\{(v_0,v_i),
(v_i,v_j), (v_j,v_l), (v_l,v_r)\}$. Let $Z_N$ and $Z_{N^U}$ be the
weights corresponding to this flow in the network $N$ and $N^U$
respectively.

Therefore,
\begin{eqnarray*}
Z_{N^U} &=& \sum_{(v_i,v_j)\in E'_N} w_{N^U}(v_i,v_j)\\
        &=& \sum_{(v_i,v_j)\in E'_N} \{\pi(v_i)-\pi(v_j)-w_{N}(v_i,v_j)\}\\
        &=& \pi(s)-\pi(t)-\sum_{(v_i,v_j)\in E'_N} w_{N}(v_i,v_j)\\
        &=& \pi(s)-\sum_{(v_i,v_j)\in E'_N} w_{N}(v_i,v_j)\;\;\; ({\rm as}\;\; \pi(t)=0)\\
        &=& \pi(s)-Z_N\\
  {\rm i.e.,}\;\; Z_N+Z_{N^U}&=&\pi(s),
\end{eqnarray*}
which is constant, that is, independent of any flow. Therefore, if
$Z_{N^U}$ is minimum then $Z_N$ is maximum. Hence the lemma
follows. \hfill $\Box$

\ccc{Let $O_{P5}$ be the solution of the problem P5. That is,
$O_{P5}$ is a set of $k$ disjoint paths and each path is a
sequence of vertices and edges ($i$-arcs and $c$-arcs). Again, let
$O_{P4}$ be the solution of the problem P4 and it is also the set
of $k$ disjoint paths containing the same sets of vertices and
edges of $O_{P5}$. Note that $k$ disjoint paths of P4 and P5 are
same, but the arc weights are different. }

 Let $O'_{P4}$ and $O'_{P5}$ \ccc{be the sets of} $i$-arcs of
$O_{P4}$ and $O_{P5}$ respectively. \ccc{Now} from the above
theorem it follows that \ccc{$O'_{P5}$} is equal to
\ccc{$O'_{P4}$}.

Combining Lemma \ref{l110} and Theorem \ref{t10}  {the following
result can be stated.}

\begin{theorem}
\label{t20} The vertices \ccc{of G} corresponding to \ccc{the}
arcs of $O'_{P5}$ are the vertices of the problem P2 for the
interval graph $G$.
\end{theorem}

In the next section, the algorithm to solve MWkC problem is
presented. Also, the time and space complexities are analyzed.

\section{The Algorithm and its Complexity}
We have discussed several properties about network $N$ in previous
sections. Also, we have shown that maximum weight $k$-flow problem
is equivalent to MWkC problem. Again, the solution of MWkC is the
solution of the problem P1. In the following, we present the major
steps of the proposed algorithm to solve maximum weight $k$-flow
problem.

\bigskip
\noindent {\sc Algorithm MWKF}\\
{\it Input:} The sorted endpoints list of intervals of an interval
graph $G$ and a positive integer $k$.\\
{\it Output:} The $k$ colour classes $H_1, H_2, \ldots, H_k$ of the graph $G$.\\
\begin{tabular}{lp{14cm}}
{\bf Step 1:} & Find all maximal cliques of the graph $G$.\\
{\bf Step 2:} & Construct a network $N$, using the technique described in
Section 5.  \\
{\bf Step 3:} & Compute the array $\pi$ using Algorithm $\Pi$. \\
{\bf Step 4:} & Convert the network $N$ to the network $N^U$ by
                changing the weights to\\
              &  $w_{N^U}(v_i,v_j)=\pi(v_i)-\pi(v_j)-w_N(v_i,v_j), \;\; \mbox{for all}\;\;  (v_i,v_j)\in
                 E_N$. \\
{\bf Step 5:} & Solve the minimum weight $k$-flow problem for the network $N^U$.\\
{\bf Step 6:} & For any $i$-arc $(u,v)\in E_N$, if $f(u,v)>0$ then put the corresponding vertex to the set $Q$. \\
{\bf Step 7:} & Distribute the vertices of $Q$ to $k$ colour classes $H_1,
H_2, \ldots, H_k$.
                The vertices of each set have same colour.\\
\end{tabular}

\noindent {\bf end MWKF}

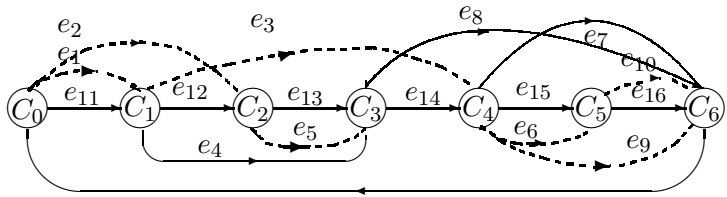
\begin{figure}[h]
\special{em:linewidth 0.4pt} \unitlength 1.00mm
\linethickness{0.4pt}
\begin{center}
\ifx\plotpoint\undefined\newsavebox{\plotpoint}\fi 
\begin{picture}(95.665,30.375)(0,0)
\put(3,12.33){\circle{5.33}} \put(18,12.33){\circle{5.33}}
\put(33,12.33){\circle{5.33}} \put(48,12.33){\circle{5.33}}
\put(63,12.33){\circle{5.33}} \put(78,12.33){\circle{5.33}}
\put(93,12.33){\circle{5.33}}
\put(48,12.33){\makebox(0,0)[cc]{$C_3$}}
\put(63,12.33){\makebox(0,0)[cc]{$C_4$}}
\put(78,12.33){\makebox(0,0)[cc]{$C_5$}}
\put(93,12.33){\makebox(0,0)[cc]{$C_6$}}
\put(3,12){\makebox(0,0)[cc]{$C_0$}}
\put(18,12.33){\makebox(0,0)[cc]{$C_1$}}
\put(33,12.33){\makebox(0,0)[cc]{$C_2$}}
\put(30.33,12.33){\vector(1,0){.2}}
\put(20.67,12.33){\line(1,0){9.66}}
\put(45.33,12.33){\vector(1,0){.2}}
\put(35.67,12.33){\line(1,0){9.66}}
\put(60.33,12.33){\vector(1,0){.2}}
\put(50.67,12.33){\line(1,0){9.66}}
\put(75.33,12.33){\vector(1,0){.2}}
\put(65.67,12.33){\line(1,0){9.66}}
\put(90.33,12.33){\vector(1,0){.2}}
\put(80.67,12.33){\line(1,0){9.66}}
\put(15.33,12.33){\vector(1,0){.2}}
\put(5.67,12.33){\line(1,0){9.66}}
\put(33.17,8.67){\oval(29.67,6.67)[b]}
\multiput(63,15)(.03319149,.0387234){47}{\line(0,1){.0387234}}
\multiput(64.56,16.82)(.03340426,.03425532){47}{\line(0,1){.03425532}}
\multiput(66.13,18.43)(.03714286,.03333333){42}{\line(1,0){.03714286}}
\multiput(67.69,19.83)(.04361111,.03333333){36}{\line(1,0){.04361111}}
\multiput(69.26,21.03)(.0523333,.0333333){30}{\line(1,0){.0523333}}
\multiput(70.83,22.03)(.0654167,.0333333){24}{\line(1,0){.0654167}}
\multiput(72.4,22.83)(.0872222,.0327778){18}{\line(1,0){.0872222}}
\multiput(73.97,23.42)(.131667,.031667){12}{\line(1,0){.131667}}
\multiput(75.55,23.8)(.261667,.031667){6}{\line(1,0){.261667}}
\put(77.12,23.99){\line(1,0){1.58}}
\multiput(78.7,23.97)(.225714,-.032857){7}{\line(1,0){.225714}}
\multiput(80.28,23.74)(.1223077,-.0330769){13}{\line(1,0){.1223077}}
\multiput(81.87,23.31)(.0831579,-.0331579){19}{\line(1,0){.0831579}}
\multiput(83.45,22.68)(.0636,-.0332){25}{\line(1,0){.0636}}
\multiput(85.04,21.85)(.0509677,-.0335484){31}{\line(1,0){.0509677}}
\multiput(86.62,20.81)(.04297297,-.03351351){37}{\line(1,0){.04297297}}
\multiput(88.21,19.57)(.03697674,-.03372093){43}{\line(1,0){.03697674}}
\multiput(89.8,18.12)(.03337209,-.03627907){86}{\line(0,-1){.03627907}}
\put(18,21){\vector(1,0){.2}} \put(16.67,21){\line(1,0){1.33}}
\put(38,7.33){\line(1,0){2.67}} \put(33.67,5.33){\vector(1,0){.2}}
\put(30,5.33){\line(1,0){3.67}} \put(71.67,7.66){\vector(1,0){.2}}
\put(69.33,7.66){\line(1,0){2.34}} \put(87,16.33){\vector(1,0){.2}}
\put(78.33,23.99){\vector(1,0){.2}} \put(76.33,23.99){\line(1,0){2}}
\put(18.33,7.33){\line(0,1){2}} \put(48,16){\line(0,-1){1}}
\put(8.67,19.33){\makebox(0,0)[cc]{$e_1$}}
\put(8.67,23){\makebox(0,0)[cc]{$e_2$}}
\put(34.33,23.66){\makebox(0,0)[cc]{$e_3$}}
\put(27.33,7){\makebox(0,0)[cc]{$e_4$}}
\put(40,9.33){\makebox(0,0)[cc]{$e_5$}}
\put(69.33,9.5){\makebox(0,0)[cc]{$e_6$}}
\put(84.34,7.66){\makebox(0,0)[cc]{$e_9$}}
\put(78.67,21.66){\makebox(0,0)[cc]{$e_7$}}
\put(84.33,18.5){\makebox(0,0)[cc]{$e_{10}$}}
\put(10.33,14){\makebox(0,0)[cc]{$e_{11}$}}
\put(24.67,14.66){\makebox(0,0)[cc]{$e_{12}$}}
\put(40,14){\makebox(0,0)[cc]{$e_{13}$}}
\put(55.67,14){\makebox(0,0)[cc]{$e_{14}$}}
\put(70.33,14.66){\makebox(0,0)[cc]{$e_{15}$}}
\put(85.67,14.33){\makebox(0,0)[cc]{$e_{16}$}} \special{em:linewidth
0.8pt}
\multiput(4.33,16)(-.03325,-.0335){40}{\line(0,-1){.0335}}
\multiput(18,14.66)(-.067,.0335){20}{\line(-1,0){.067}}
\multiput(33.66,8.66)(-.0333333,.0333333){30}{\line(0,1){.0333333}}
\put(48,9.5){\oval(90,16.33)[b]} \put(46.67,1.33){\vector(-1,0){.2}}
\put(49,1.33){\line(-1,0){2.33}} \thicklines
\multiput(3.18,14.93)(.0446429,.0334821){16}{\line(1,0){.0446429}}
\multiput(4.608,16.001)(.0446429,.0334821){16}{\line(1,0){.0446429}}
\multiput(6.037,17.073)(.0446429,.0334821){16}{\line(1,0){.0446429}}
\multiput(7.465,18.144)(.0446429,.0334821){16}{\line(1,0){.0446429}}
\multiput(17.93,21.18)(.20833,-.03125){4}{\line(1,0){.20833}}
\multiput(19.596,20.93)(.20833,-.03125){4}{\line(1,0){.20833}}
\multiput(21.263,20.68)(.20833,-.03125){4}{\line(1,0){.20833}}
\multiput(22.93,20.43)(.05625,-.03125){10}{\line(1,0){.05625}}
\multiput(24.055,19.805)(.05625,-.03125){10}{\line(1,0){.05625}}
\multiput(25.18,19.18)(.05,-.033333){12}{\line(1,0){.05}}
\multiput(26.38,18.38)(.05,-.033333){12}{\line(1,0){.05}}
\multiput(27.58,17.58)(.05,-.033333){12}{\line(1,0){.05}}
\multiput(28.18,17.18)(.0366667,-.0333333){15}{\line(1,0){.0366667}}
\multiput(29.28,16.18)(.0366667,-.0333333){15}{\line(1,0){.0366667}}
\multiput(30.38,15.18)(.0366667,-.0333333){15}{\line(1,0){.0366667}}
\multiput(18.93,14.93)(.084375,.03125){10}{\line(1,0){.084375}}
\multiput(20.617,15.555)(.084375,.03125){10}{\line(1,0){.084375}}
\multiput(22.305,16.18)(.084375,.03125){10}{\line(1,0){.084375}}
\multiput(23.992,16.805)(.084375,.03125){10}{\line(1,0){.084375}}
\multiput(25.68,17.43)(.16,.03){5}{\line(1,0){.16}}
\multiput(27.28,17.73)(.16,.03){5}{\line(1,0){.16}}
\multiput(28.88,18.03)(.16,.03){5}{\line(1,0){.16}}
\multiput(29.68,18.18)(.15625,.03125){6}{\line(1,0){.15625}}
\multiput(31.555,18.555)(.15625,.03125){6}{\line(1,0){.15625}}
\multiput(33.43,18.93)(.16,.03){5}{\line(1,0){.16}}
\multiput(35.03,19.23)(.16,.03){5}{\line(1,0){.16}}
\multiput(36.63,19.53)(.16,.03){5}{\line(1,0){.16}}
\multiput(37.43,19.68)(.23438,.03125){4}{\line(1,0){.23438}}
\multiput(39.305,19.93)(.23438,.03125){4}{\line(1,0){.23438}}
\put(41.18,20.18){\line(1,0){.875}}
\put(42.93,20.18){\line(1,0){.875}}
\put(44.68,20.18){\line(1,0){.8}} \put(46.28,20.28){\line(1,0){.8}}
\put(47.88,20.38){\line(1,0){.8}}
\multiput(48.68,20.43)(.1875,-.03125){4}{\line(1,0){.1875}}
\multiput(50.18,20.18)(.1875,-.03125){4}{\line(1,0){.1875}}
\multiput(51.68,19.93)(.1,-.033333){6}{\line(1,0){.1}}
\multiput(52.88,19.53)(.1,-.033333){6}{\line(1,0){.1}}
\multiput(54.08,19.13)(.1,-.033333){6}{\line(1,0){.1}}
\multiput(54.68,18.93)(.108333,-.033333){6}{\line(1,0){.108333}}
\multiput(55.98,18.53)(.108333,-.033333){6}{\line(1,0){.108333}}
\multiput(57.28,18.13)(.108333,-.033333){6}{\line(1,0){.108333}}
\multiput(57.93,17.93)(.0535714,-.0327381){14}{\line(1,0){.0535714}}
\multiput(59.43,17.013)(.0535714,-.0327381){14}{\line(1,0){.0535714}}
\multiput(60.93,16.096)(.0535714,-.0327381){14}{\line(1,0){.0535714}}
\multiput(32.93,9.68)(.0336538,-.0384615){13}{\line(0,-1){.0384615}}
\multiput(33.805,8.68)(.0336538,-.0384615){13}{\line(0,-1){.0384615}}
\multiput(34.68,7.68)(.166667,-.033333){5}{\line(1,0){.166667}}
\multiput(36.346,7.346)(.166667,-.033333){5}{\line(1,0){.166667}}
\put(37.18,7.18){\line(1,0){.8333}}
\put(38.846,7.18){\line(1,0){.8333}}
\put(39.68,7.18){\line(1,0){.8}} \put(41.28,7.28){\line(1,0){.8}}
\put(42.88,7.38){\line(1,0){.8}}
\multiput(43.68,7.43)(.085938,.03125){8}{\line(1,0){.085938}}
\multiput(45.055,7.93)(.085938,.03125){8}{\line(1,0){.085938}}
\multiput(46.43,8.43)(.0384615,.0320513){13}{\line(1,0){.0384615}}
\multiput(47.43,9.263)(.0384615,.0320513){13}{\line(1,0){.0384615}}
\put(63.18,9.68){\line(0,-1){.125}}
\multiput(63.18,9.93)(.0528846,-.0336538){13}{\line(1,0){.0528846}}
\multiput(64.555,9.055)(.0528846,-.0336538){13}{\line(1,0){.0528846}}
\multiput(65.93,8.18)(.133333,-.033333){5}{\line(1,0){.133333}}
\multiput(67.263,7.846)(.133333,-.033333){5}{\line(1,0){.133333}}
\multiput(67.93,7.68)(.21875,-.03125){4}{\line(1,0){.21875}}
\put(69.68,7.43){\line(1,0){.75}} \put(71.18,7.555){\line(1,0){.75}}
\multiput(72.68,7.68)(.135417,.03125){6}{\line(1,0){.135417}}
\multiput(74.305,8.055)(.135417,.03125){6}{\line(1,0){.135417}}
\multiput(75.93,8.43)(.0625,.03125){8}{\line(1,0){.0625}}
\multiput(76.93,8.93)(.0625,.03125){8}{\line(1,0){.0625}}
\multiput(63.43,9.93)(.041667,-.033333){12}{\line(1,0){.041667}}
\multiput(64.43,9.13)(.041667,-.033333){12}{\line(1,0){.041667}}
\multiput(65.43,8.33)(.041667,-.033333){12}{\line(1,0){.041667}}
\multiput(65.93,7.93)(.0625,-.03125){10}{\line(1,0){.0625}}
\multiput(67.18,7.305)(.0625,-.03125){10}{\line(1,0){.0625}}
\multiput(68.43,6.68)(.108333,-.033333){6}{\line(1,0){.108333}}
\multiput(69.73,6.28)(.108333,-.033333){6}{\line(1,0){.108333}}
\multiput(71.03,5.88)(.108333,-.033333){6}{\line(1,0){.108333}}
\multiput(71.68,5.68)(.125,-.03125){6}{\line(1,0){.125}}
\multiput(73.18,5.305)(.125,-.03125){6}{\line(1,0){.125}}
\put(74.68,4.93){\line(1,0){.8333}}
\put(76.346,4.763){\line(1,0){.8333}}
\multiput(81.68,4.68)(.107143,.029762){7}{\line(1,0){.107143}}
\multiput(83.18,5.096)(.107143,.029762){7}{\line(1,0){.107143}}
\multiput(84.68,5.513)(.107143,.029762){7}{\line(1,0){.107143}}
\multiput(86.18,5.93)(.054545,.031818){11}{\line(1,0){.054545}}
\multiput(87.38,6.63)(.054545,.031818){11}{\line(1,0){.054545}}
\multiput(88.58,7.33)(.054545,.031818){11}{\line(1,0){.054545}}
\multiput(89.18,7.68)(.0321429,.0357143){14}{\line(0,1){.0357143}}
\multiput(90.08,8.68)(.0321429,.0357143){14}{\line(0,1){.0357143}}
\multiput(90.98,9.68)(.0321429,.0357143){14}{\line(0,1){.0357143}}
\multiput(79.68,14.18)(.06875,.03125){10}{\line(1,0){.06875}}
\multiput(81.055,14.805)(.06875,.03125){10}{\line(1,0){.06875}}
\multiput(82.43,15.43)(.1875,.03125){4}{\line(1,0){.1875}}
\multiput(83.93,15.68)(.1875,.03125){4}{\line(1,0){.1875}}
\multiput(88.68,16.18)(.057292,-.03125){12}{\line(1,0){.057292}}
\multiput(90.055,15.43)(.057292,-.03125){12}{\line(1,0){.057292}}
\multiput(7.43,18.18)(.054167,.033333){12}{\line(1,0){.054167}}
\multiput(8.73,18.98)(.054167,.033333){12}{\line(1,0){.054167}}
\multiput(10.03,19.78)(.054167,.033333){12}{\line(1,0){.054167}}
\multiput(10.68,20.18)(.135417,.03125){6}{\line(1,0){.135417}}
\multiput(12.305,20.555)(.135417,.03125){6}{\line(1,0){.135417}}
\put(13.93,20.93){\line(1,0){.75}}
\put(15.43,21.055){\line(1,0){.75}}
\put(38.5,19.75){\vector(4,1){.07}}\put(36.43,19.43){\line(1,0){.6667}}
\put(37.763,19.596){\line(1,0){.6667}}
\put(80.5,4.75){\vector(1,0){.07}}\put(78.18,4.43){\line(1,0){.75}}
\put(79.68,4.596){\line(1,0){.75}}
\put(40,7.25){\vector(1,0){.07}}\put(37.68,7.18){\line(1,0){.75}}
\put(39.18,7.18){\line(1,0){.75}}
\multiput(3.18,14.68)(.041667,.03125){12}{\line(1,0){.041667}}
\multiput(4.18,15.43)(.041667,.03125){12}{\line(1,0){.041667}}
\multiput(5.18,16.18)(.09375,.03125){8}{\line(1,0){.09375}}
\multiput(6.68,16.68)(.09375,.03125){8}{\line(1,0){.09375}}
\multiput(7.43,16.93)(.166667,.033333){5}{\line(1,0){.166667}}
\multiput(9.096,17.263)(.166667,.033333){5}{\line(1,0){.166667}}
\multiput(13.68,16.93)(.09375,-.03125){8}{\line(1,0){.09375}}
\multiput(15.18,16.43)(.072917,-.03125){12}{\line(1,0){.072917}}
\multiput(16.93,15.68)(.03125,-.03125){12}{\line(0,-1){.03125}}
\put(12.25,17.25){\vector(1,0){.07}}\put(10.93,17.18){\line(1,0){.625}}
\thinlines
\put(92.75,15){\vector(2,-1){.07}}\qbezier(48,15.25)(60.125,30.375)(92.75,15)
\put(62,24.75){\makebox(0,0)[cc]{$e_8$}}
\put(64.25,22.5){\vector(1,0){.07}}\put(62.18,22.43){\line(1,0){.6667}}
\put(63.513,22.43){\line(1,0){.6667}}
\end{picture}
\end{center}
\caption{\label{f30} {Dotted arcs} are the solution of the MW2C
problem of the graph of Figure \ref{f10}. }
\end{figure}

\bigskip
In Figure \ref{f30}, the {dotted} arcs are the arcs of the minimum
weight 2-flow for the network $N^U$.

The arcs of the minimum weight 2-flow of the network $N^U$ are $e_1,
e_2, e_3, e_5, e_6, e_9, e_{10}$ and $\{1, 2, 5, 3, 6, 9, 10\}$ are
the vertices corresponding to the arcs $e_1, e_2, e_3, e_5, e_6,
e_9, e_{10}$. The set $Q$ after Step 6 of Algorithm MWKF is $\{1, 2,
3, 5, 6, 9, 10\}$. The total weight (maximum) of 2-flow is 34 (the
sum of weights of the vertices of $Q$) and the 2 flows are $\{1, 5,
6, 10\}$ and $\{2, 3, 9\}$. The other solutions are $\{1, 5, 9\}$,
$\{2, 3, 6, 10\}$.

\begin{theorem}
\label{t30} The running time of Algorithm MWKF is $O(kn\sqrt{\log
c}+m)$, where $n$ and $m$ represent respectively the number of
vertices and edges and $c$ is the weight of the longest path of the
interval graph.
\end{theorem}

\noindent {\bf Proof.} All maximal cliques of an interval graph can
be computed in $O(n+\gamma)$ time, where $\gamma$ is the total size
of all maximal cliques \cite{pal95}. But, $\gamma$ is of order
$O(n+m)$ \cite{gol80}. The network $N$ can be constructed using
$O(n)$ time. To compute the array $\pi$ and the weight of each arc
of the network $N^U$ take $O(n)$ time. The minimum weight $k$-flow
problem of $N^U$ can be solved using the algorithm of Edmonds and
Karp in time $O(k\times \mbox{(complexity of shortest path
problem)})$. An $O(m+n \sqrt{\log c})$ time algorithm \cite{ahu90}
is available to solve the shortest path problem on general graph
with $n$ vertices and $m$ edges, where cost of each arc is
non-negative integer number bounded by $c$. In $N^U$, the weight of
each arc is bounded by $\pi(s)$ $(\le c)$, if $c$ is the largest
weight of a path in the given interval graph. Thus, since there are
$O(n)$ arcs in $N^U$, the algorithm requires $O(kn\sqrt{\log c}+m)$
operations to solve the $k$-flow problem. The Step 6 requires only
$O(n)$ time. The Step 7 can be computed using $O(kn)$ time.
Therefore, the total time complexity is $O(kn \sqrt{\log c}+m)$.
\hfill $\Box$


The upper bound of the weights of arcs in $N$ or $N^U$ is
$\pi(s)$. If $M$ be the upper bound of weights of the vertices of
the given interval graph then the upper bound of $c$ is $nM$. From
this analogy we can draw the following conclusion.

\begin{theorem}
\label{t40} The running time of Algorithm MWKF is $O(k M n^2)$,
where $n$ and $M$ represent respectively the number of vertices
and the upper bound of the weights of the vertices of the interval
graph.
\end{theorem}

{From this theorem one can conclude that the maximum weight
$k$-colourable subgraph problem on interval graph can be solved
using $O(k M n^2)$ time, where $n$ and $M$ represent respectively
the number of vertices and the upper bound of weights of the
vertices of the interval graph.}

At the beginning of this article, it is mentioned that a daily
life problem is considered in this paper. This practical problem
is solved by using the concept of graph theory. In the next
section, a numerical example is considered for illustration.

\section{\ccc{Numerical Illustration}}
We shall use some standard notations to name the television
programme for convenience of solving the problem. The name of the
programmes of various channels and the number of viewers (taken as
the weight of the interval) are written in parentheses. The first
number of the parentheses represents the name of the programme and
the second number tells about weight or strength of viewer in lakh.

Here we consider three channels and try to find out a solution of
the problem. The channels we consider are National Geographic,
Discovery, and AXN. Some programmes of National Geographic are as
follows: 8:00--9:00 a.m. India Diaries: Keeping faith (N$_1$, 5);
9:00--10:00 a.m. Reel people : Through these eyes (N$_2$, 7);
10:00--10:30 a.m. Mission Everest (N$_3$, 8); 10:30--11:00 a.m.
Nick's Quest (N$_4$, 3); 11:00--11:30 p.m. Wild orphan (N$_5$, 4);
11:30--12:00 noon Myths and logic of shoolin kung fu (N$_6$, 5);
12:00--1:00 p.m. Adventure atarts here with Toyota (N$_7$, 4) and
so on.

\setlength{\tabcolsep}{.2mm}
\begin{table}
\begin{tabular}{l|cccccccc}
\hline Time & 8:00-9:00 & 9:00-10:00 &10:00-10:30 &10:30-11:00
&11:00-11:30
&11:30-12:00 &12:00-13:00\\
\hline Short name & N$_1$ &  N$_2$ &    N$_3$  &  N$_4$  &  N$_5$
  &N$_6$ & N$_7$\\
Viewers & 5         &  7         &    8       &  3         &   4
&
5 &  4\\
\hline Time & 13:00-13:30 & 13:30-14:00 &14:00-14:30 &14:30-15:30
&15:30-16:00
&16:00-17:00 &17:00-17:30\\
\hline Short name &   N$_8$   &  N$_9$ &   N$_{10}$  &  N$_{11}$ &
N$_{12}$   & N$_{13}$ & N$_{14}$       \\
Viewers&  7         &  8          &    2       &  4         &   5
&
6         &  4\\
\hline Time & 17:30-18:00 & 18:00-18:30 &18:30-19:00 &19:00-19:30
&19:30-20:30
&20:30-21:30 &21:30-22:00\\
\hline Short name &   N$_{15}$ &  N$_{16}$&    N$_{17}$ &
N$_{18}$       &
 N$_{19}$   & N$_{20}$ & N$_{21}$       \\
Viewers &  3         &  1          &    2       &  3         &   6
&
 1         &  4\\
\hline Time & 22:00-22:30 & 22:30-23:00 &            &
&            &
   &   \\
\hline Short name &  N$_{22}$&  N$_{23}$        &            &
&
  &            &   \\
Viewers &  3         &  1          &            &            &
&
   &   \\
\hline
\end{tabular}
\caption{\label{tab20} Programmes of the channel National
Geography.}
\end{table}

The programmes of Discovery channel are as follows:\\
8:00--9:00 a.m. Terra X (D$_1$, 5); 9:00--10:00 a.m. Tunk yard war
kids (D$_2$, 7); 10:00--10:30 a.m. Discover India (D$_3$, 8);
10:30--11:00 a.m. Real kids real adventure (D$_4$, 3);
11:00--11:30 a.m. Eccentriks (D$_5$, 4); 11:30--12:00 noon Djuma :
South Africa (D$_6$, 5); 12:00--12:30 p.m. Wedding Story (D$_7$,
4), etc.

\begin{table}
\begin{tabular}{l|cccccccc}
\hline Time & 8:00-9:00 & 9:00-10:00 &10:00-10:30 &10:30-11:00
&11:00-11:30
&11:30-12:00 &12:00-12:30\\
\hline
Short name & $D_1$ & $D_2$ &$D_3$&  $D_4$ &  $D_5$ & $D_6$ & $D_7$ \\
Viewers & 5         &  7         &    8       &  3         &   4
&
5  &  4\\
\hline Time & 12:30-13:00 & 13:00-14:00 &14:00-15:00 &15:00-16:00
&16:00-17:00
&17:00-18:00 &18:00-18:30\\
\hline Short name& $D_8$& $D_9$ & $D_{10}$ &$D_{11}$& $D_{12}$ &
$D_{13}$ &
$D_{14}$\\
Viewers &  7         &  8          &    2       &  4         &   5
&
 6    &  4\\
\hline Time & 18:30-19:00 & 19:00-20:00 &20:00-21:00 &21:00-22:00
&22:00-23:00 &
   &           \\
\hline
Short name&   $D_{15}$&  $D_{16}$&$D_{17}$ &  $D_{18}$& $D_{19}$ & &\\
Viewers &  2         &  3          &    1       &  4         &   2
&
   &   \\
\hline
\end{tabular}
\caption{\label{tab30} Programmes of the channel Discovery.}
\end{table}

The programmes of AXN with time and number of viewers are such as\\
7:00--8:00 a.m. Relic Hunter (A$_1$,3); 8:00--9:00 a.m. Now see
this (A$_2$,1); 9:00--10:00 a.m. Guiness World records (A$_3$,2);
10:00--11:00 a.m. Ripley's Believe it or Not (A$_4$,3), etc.

The tabular representation of the programmes of different channels
are shown in the tables \ref{tab20}, \ref{tab30} and \ref{tab40}.

\begin{table}
\begin{tabular}{l|cccccccc}
\hline Time & 7:00-8:00 & 8:00-9:00 &9 :00-10:00 &10:00-11:00
&11:00-12:00
&12:00-15:00 &15:00-17:30\\
\hline
Short name & $A_1$&$A_2$ & $A_3$ &  $A_4$ & $A_5$ & $A_6$ & $A_7$\\
Viewers & 3         &  1         &    2       &  3         &   2
&
5  &  4\\
\hline Time & 17:30-18:00 & 18:00-19:00 &19:00-20:00 &20:00-22:30
&22:30-24:00 &
&\\
\hline
Short name & $A_8$& $A_9$&$A_{10}$&  $A_{11}$ &  $A_{12}$ & &\\
Viewers&  3         &  4          &    2       &  4         &   2
&
&\\
\hline
\end{tabular}
\caption{\label{tab40} Programmes of the channel AXN.}
\end{table}

The combined {programme slots} of three channels are shown in the
Figure \ref{f40}.

{\scriptsize
\begin{figure}
\begin{center}
\special{em:linewidth 0.4pt} \unitlength 1.20mm
\linethickness{1.4pt}
\begin{picture}(95.00,39.00)
\emline{5.00}{3.67}{1}{5.00}{6.33}{2}
\emline{10.00}{6.33}{3}{10.00}{3.67}{4}
\emline{15.00}{3.67}{5}{15.00}{6.33}{6}
\emline{20.00}{6.33}{7}{20.00}{3.67}{8}
\emline{25.00}{6.33}{9}{25.00}{3.67}{10}
\emline{30.00}{3.67}{11}{30.00}{6.33}{12}
\emline{35.00}{6.33}{13}{35.00}{3.67}{14}
\emline{40.00}{3.67}{15}{40.00}{6.00}{16}
\emline{50.00}{6.33}{17}{50.00}{3.67}{18}
\emline{55.00}{3.67}{19}{55.00}{6.33}{20}
\emline{60.00}{6.33}{21}{60.00}{3.67}{22}
\emline{65.00}{3.67}{23}{65.00}{6.33}{24}
\emline{70.00}{6.33}{25}{70.00}{3.67}{26}
\emline{75.00}{3.67}{27}{75.00}{6.33}{28}
\emline{80.00}{6.33}{29}{80.00}{3.67}{30}
\emline{85.00}{3.67}{31}{85.00}{6.33}{32}
\put(5.00,1.33){\makebox(0,0)[cc]{7}}
\put(10.00,1.33){\makebox(0,0)[cc]{8}}
\put(15.00,1.33){\makebox(0,0)[cc]{9}}
\put(20.00,1.33){\makebox(0,0)[cc]{10}}
\put(25.00,1.33){\makebox(0,0)[cc]{11}}
\put(30.00,1.33){\makebox(0,0)[cc]{12}}
\put(35.00,1.33){\makebox(0,0)[cc]{13}}
\put(40.00,1.33){\makebox(0,0)[cc]{14}}
\put(45.00,1.33){\makebox(0,0)[cc]{15}}
\put(50.00,1.33){\makebox(0,0)[cc]{16}}
\put(55.00,1.33){\makebox(0,0)[cc]{17}}
\put(60.00,1.33){\makebox(0,0)[cc]{18}}
\put(65.00,1.33){\makebox(0,0)[cc]{19}}
\put(70.00,1.33){\makebox(0,0)[cc]{20}}
\put(75.00,1.33){\makebox(0,0)[cc]{21}}
\put(80.00,1.33){\makebox(0,0)[cc]{22}}
\put(85.00,1.33){\makebox(0,0)[cc]{23}}
\put(90.00,1.33){\makebox(0,0)[cc]{24}}
\emline{90.00}{6.33}{33}{90.00}{3.67}{34}
\emline{95.00}{5.00}{35}{0.00}{5.00}{36}
\emline{45.00}{3.67}{37}{45.00}{6.33}{38}
\emline{10.00}{9.67}{39}{15.00}{9.67}{40}
\emline{20.00}{9.67}{41}{22.33}{9.67}{42}
\emline{15.00}{14.00}{43}{20.00}{14.00}{44}
\emline{22.67}{14.00}{45}{25.00}{14.00}{46}
\emline{25.00}{9.67}{47}{27.67}{9.67}{48}
\emline{27.67}{14.00}{49}{30.00}{14.00}{50}
\emline{30.00}{9.67}{51}{35.00}{9.67}{52}
\emline{35.00}{14.00}{53}{37.33}{14.00}{54}
\emline{37.33}{9.67}{55}{40.00}{9.67}{56}
\emline{40.00}{14.00}{57}{42.33}{14.00}{58}
\emline{42.33}{9.67}{59}{47.33}{9.67}{60}
\emline{47.33}{14.00}{61}{50.00}{14.00}{62}
\emline{50.00}{9.67}{63}{55.00}{9.67}{64}
\emline{55.00}{14.00}{65}{57.33}{14.00}{66}
\emline{57.33}{9.67}{67}{60.00}{9.67}{68}
\emline{60.00}{14.00}{69}{62.67}{14.00}{70}
\emline{62.67}{9.67}{71}{65.00}{9.67}{72}
\emline{65.00}{14.00}{73}{67.67}{14.00}{74}
\emline{67.67}{9.67}{75}{70.00}{9.67}{76}
\emline{70.00}{9.67}{77}{72.33}{9.67}{78}
\emline{72.33}{14.00}{79}{77.33}{14.00}{80}
\emline{77.33}{9.67}{81}{80.00}{9.67}{82}
\emline{80.00}{14.00}{83}{82.33}{14.00}{84}
\emline{82.33}{9.67}{85}{85.00}{9.67}{86}
\put(12.67,11.00){\makebox(0,0)[cc]{$N_1$}}
\put(21.00,11.33){\makebox(0,0)[cc]{$N_3$}}
\put(26.00,11.33){\makebox(0,0)[cc]{$N_5$}}
\put(32.33,11.33){\makebox(0,0)[cc]{$N_7$}}
\put(38.67,11.33){\makebox(0,0)[cc]{$N_9$}}
\put(44.67,11.33){\makebox(0,0)[cc]{$N_{11}$}}
\put(52.33,11.33){\makebox(0,0)[cc]{$N_{13}$}}
\put(58.67,11.33){\makebox(0,0)[cc]{$N_{15}$}}
\put(63.67,11.33){\makebox(0,0)[cc]{$N_{17}$}}
\put(69.67,11.33){\makebox(0,0)[cc]{$N_{19}$}}
\put(78.67,11.33){\makebox(0,0)[cc]{$N_{21}$}}
\put(83.67,11.33){\makebox(0,0)[cc]{$N_{23}$}}
\put(17.33,15.67){\makebox(0,0)[cc]{$N_2$}}
\put(24.00,15.67){\makebox(0,0)[cc]{$N_4$}}
\put(28.67,15.67){\makebox(0,0)[cc]{$N_6$}}
\put(36.00,15.67){\makebox(0,0)[cc]{$N_8$}}
\put(41.33,15.67){\makebox(0,0)[cc]{$N_{10}$}}
\put(48.67,15.67){\makebox(0,0)[cc]{$N_{12}$}}
\put(56.33,15.67){\makebox(0,0)[cc]{$N_{14}$}}
\put(61.33,15.67){\makebox(0,0)[cc]{$N_{16}$}}
\put(66.33,15.67){\makebox(0,0)[cc]{$N_{18}$}}
\put(74.67,15.67){\makebox(0,0)[cc]{$N_{20}$}}
\put(81.00,15.67){\makebox(0,0)[cc]{$N_{22}$}}
\emline{10.00}{20.33}{87}{15.00}{20.33}{88}
\emline{15.00}{25.00}{89}{20.00}{25.00}{90}
\emline{20.00}{20.33}{91}{22.33}{20.33}{92}
\emline{22.33}{25.00}{93}{25.00}{25.00}{94}
\emline{25.00}{20.33}{95}{27.33}{20.33}{96}
\emline{27.33}{25.00}{97}{30.00}{25.00}{98}
\emline{30.00}{20.33}{99}{32.33}{20.33}{100}
\emline{32.33}{25.00}{101}{35.00}{25.00}{102}
\emline{35.00}{20.33}{103}{40.00}{20.33}{104}
\emline{40.00}{25.00}{105}{45.00}{25.00}{106}
\emline{45.00}{20.33}{107}{50.00}{20.33}{108}
\emline{50.00}{25.00}{109}{55.00}{25.00}{110}
\emline{55.00}{20.33}{111}{60.00}{20.33}{112}
\emline{60.00}{25.00}{113}{62.33}{25.00}{114}
\emline{62.33}{20.33}{115}{65.00}{20.33}{116}
\emline{65.00}{25.00}{117}{70.00}{25.00}{118}
\emline{70.00}{20.33}{119}{75.00}{20.33}{120}
\emline{75.00}{25.00}{121}{80.00}{25.00}{122}
\emline{80.00}{20.33}{123}{85.00}{20.33}{124}
\put(12.33,22.00){\makebox(0,0)[cc]{$D_1$}}
\put(21.33,22.00){\makebox(0,0)[cc]{$D_3$}}
\put(26.33,22.00){\makebox(0,0)[cc]{$D_5$}}
\put(31.00,22.00){\makebox(0,0)[cc]{$D_7$}}
\put(37.67,22.00){\makebox(0,0)[cc]{$D_9$}}
\put(47.67,22.00){\makebox(0,0)[cc]{$D_{11}$}}
\put(57.33,22.00){\makebox(0,0)[cc]{$D_{13}$}}
\put(63.67,22.00){\makebox(0,0)[cc]{$D_{15}$}}
\put(72.33,22.00){\makebox(0,0)[cc]{$D_{17}$}}
\put(82.33,22.00){\makebox(0,0)[cc]{$D_{19}$}}
\put(17.67,27.00){\makebox(0,0)[cc]{$D_2$}}
\put(23.67,27.00){\makebox(0,0)[cc]{$D_4$}}
\put(28.67,27.00){\makebox(0,0)[cc]{$D_6$}}
\put(33.67,27.00){\makebox(0,0)[cc]{$D_8$}}
\put(42.33,27.00){\makebox(0,0)[cc]{$D_{10}$}}
\put(52.67,27.00){\makebox(0,0)[cc]{$D_{12}$}}
\put(61.00,27.00){\makebox(0,0)[cc]{$D_{14}$}}
\put(67.33,27.00){\makebox(0,0)[cc]{$D_{16}$}}
\put(77.33,27.00){\makebox(0,0)[cc]{$D_{18}$}}
\emline{5.00}{32.00}{125}{10.00}{32.00}{126}
\emline{10.00}{36.67}{127}{15.00}{36.67}{128}
\emline{15.00}{32.00}{129}{20.00}{32.00}{130}
\emline{20.00}{36.67}{131}{25.00}{36.67}{132}
\emline{25.00}{32.00}{133}{30.00}{32.00}{134}
\emline{30.00}{36.67}{135}{45.00}{36.67}{136}
\emline{45.00}{32.00}{137}{57.33}{32.00}{138}
\emline{57.33}{36.67}{139}{60.00}{36.67}{140}
\emline{60.00}{32.00}{141}{65.00}{32.00}{142}
\emline{65.00}{36.67}{143}{70.00}{36.67}{144}
\emline{70.00}{32.00}{145}{82.33}{32.00}{146}
\emline{82.00}{36.67}{147}{90.00}{36.67}{148}
\put(7.33,33.67){\makebox(0,0)[cc]{$A_1$}}
\put(17.67,33.67){\makebox(0,0)[cc]{$A_3$}}
\put(27.33,33.67){\makebox(0,0)[cc]{$A_5$}}
\put(51.67,33.67){\makebox(0,0)[cc]{$A_7$}}
\put(62.33,33.67){\makebox(0,0)[cc]{$A_9$}}
\put(76.00,33.67){\makebox(0,0)[cc]{$A_{11}$}}
\put(12.67,39.00){\makebox(0,0)[cc]{$A_2$}}
\put(22.33,39.00){\makebox(0,0)[cc]{$A_4$}}
\put(37.33,39.00){\makebox(0,0)[cc]{$A_6$}}
\put(58.67,39.00){\makebox(0,0)[cc]{$A_8$}}
\put(67.67,39.00){\makebox(0,0)[cc]{$A_{10}$}}
\put(86.00,39.00){\makebox(0,0)[cc]{$A_{12}$}}
\end{picture}
\end{center}
\caption{\label{f40} The set of intervals for the three channels.
The {programme slots} are shown in the tables \ref{tab20},
\ref{tab30}, \ref{tab40}. }
\end{figure}
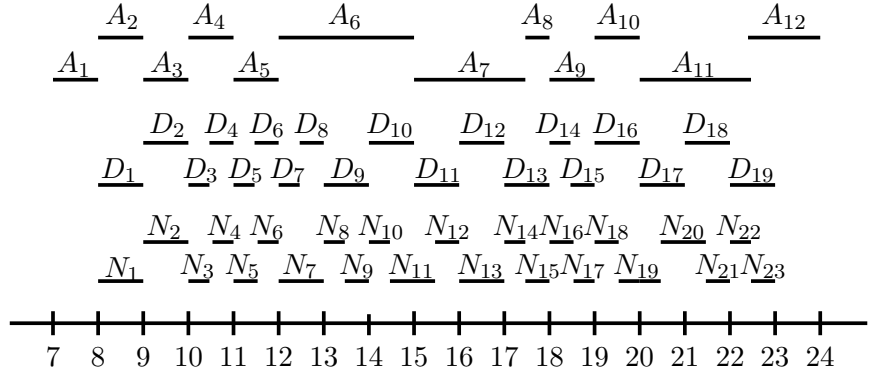
}

\newpage
The corresponding interval graph is shown in Figure \ref{f50}.

{\scriptsize
\begin{figure}
\begin{center}
\special{em:linewidth 0.4pt} \unitlength 1.00mm
\linethickness{0.4pt}
\begin{picture}(87.33,33.67)
\put(18.34,16.34){\circle{4.67}} \put(33.34,16.34){\circle{4.67}}
\put(46.34,16.00){\circle{4.67}} \put(58.34,16.34){\circle{4.67}}
\put(71.34,16.34){\circle{4.67}} \put(85.00,16.34){\circle{4.67}}
\put(3.34,31.34){\circle{4.67}} \put(18.34,31.34){\circle{4.67}}
\put(34.00,31.00){\circle{4.67}} \put(46.00,31.34){\circle{4.67}}
\put(72.00,31.00){\circle{4.67}} \put(18.34,2.34){\circle{4.67}}
\put(33.34,2.34){\circle{4.67}} \put(46.34,2.00){\circle{4.67}}
\put(58.34,2.34){\circle{4.67}} \put(71.34,2.34){\circle{4.67}}
\put(85.00,2.34){\circle{4.67}}
\put(3.34,31.34){\makebox(0,0)[cc]{$A_1$}}
\put(18.34,31.34){\makebox(0,0)[cc]{$A_2$}}
\put(34.00,31.34){\makebox(0,0)[cc]{$A_3$}}
\put(46.00,31.34){\makebox(0,0)[cc]{$A_4$}}
\put(72.34,31.00){\makebox(0,0)[cc]{$A_5$}}
\put(18.34,16.34){\makebox(0,0)[cc]{$D_1$}}
\put(33.34,16.34){\makebox(0,0)[cc]{$D_2$}}
\put(46.34,16.34){\makebox(0,0)[cc]{$D_3$}}
\put(58.34,16.34){\makebox(0,0)[cc]{$D_4$}}
\put(71.34,16.34){\makebox(0,0)[cc]{$D_5$}}
\put(85.00,16.34){\makebox(0,0)[cc]{$D_6$}}
\put(18.34,2.34){\makebox(0,0)[cc]{$N_1$}}
\put(33.67,2.34){\makebox(0,0)[cc]{$N_2$}}
\put(46.34,2.34){\makebox(0,0)[cc]{$N_3$}}
\put(58.34,2.34){\makebox(0,0)[cc]{$N_4$}}
\put(71.34,2.34){\makebox(0,0)[cc]{$N_5$}}
\put(85.00,2.34){\makebox(0,0)[cc]{$N_6$}}
\emline{18.00}{29.00}{1}{18.00}{18.67}{2}
\emline{18.00}{14.00}{3}{18.00}{4.67}{4}
\emline{16.00}{2.00}{5}{15.13}{4.34}{6}
\emline{15.13}{4.34}{7}{14.41}{6.68}{8}
\emline{14.41}{6.68}{9}{13.84}{9.01}{10}
\emline{13.84}{9.01}{11}{13.42}{11.34}{12}
\emline{13.42}{11.34}{13}{13.14}{13.67}{14}
\emline{13.14}{13.67}{15}{13.01}{15.99}{16}
\emline{13.01}{15.99}{17}{13.02}{18.30}{18}
\emline{13.02}{18.30}{19}{13.18}{20.62}{20}
\emline{13.18}{20.62}{21}{13.49}{22.93}{22}
\emline{13.49}{22.93}{23}{13.95}{25.23}{24}
\emline{13.95}{25.23}{25}{14.55}{27.53}{26}
\emline{14.55}{27.53}{27}{16.00}{31.67}{28}
\emline{31.67}{31.00}{29}{30.65}{28.76}{30}
\emline{30.65}{28.76}{31}{29.80}{26.51}{32}
\emline{29.80}{26.51}{33}{29.12}{24.26}{34}
\emline{29.12}{24.26}{35}{28.59}{22.01}{36}
\emline{28.59}{22.01}{37}{28.23}{19.75}{38}
\emline{28.23}{19.75}{39}{28.03}{17.49}{40}
\emline{28.03}{17.49}{41}{28.00}{15.22}{42}
\emline{28.00}{15.22}{43}{28.12}{12.95}{44}
\emline{28.12}{12.95}{45}{28.41}{10.68}{46}
\emline{28.41}{10.68}{47}{28.87}{8.40}{48}
\emline{28.87}{8.40}{49}{29.48}{6.12}{50}
\emline{29.48}{6.12}{51}{31.00}{2.00}{52}
\emline{33.67}{28.67}{53}{33.67}{18.67}{54}
\emline{33.34}{14.00}{55}{33.34}{4.67}{56}
\emline{45.67}{29.00}{57}{46.33}{18.33}{58}
\emline{45.67}{29.00}{59}{58.33}{18.67}{60}
\emline{45.67}{29.00}{61}{58.33}{4.67}{62}
\emline{44.00}{2.00}{63}{43.18}{4.15}{64}
\emline{43.18}{4.15}{65}{42.50}{6.33}{66}
\emline{42.50}{6.33}{67}{41.95}{8.53}{68}
\emline{41.95}{8.53}{69}{41.53}{10.77}{70}
\emline{41.53}{10.77}{71}{41.25}{13.03}{72}
\emline{41.25}{13.03}{73}{41.10}{15.33}{74}
\emline{41.10}{15.33}{75}{41.09}{17.65}{76}
\emline{41.09}{17.65}{77}{41.21}{19.99}{78}
\emline{41.21}{19.99}{79}{41.47}{22.37}{80}
\emline{41.47}{22.37}{81}{41.86}{24.78}{82}
\emline{41.86}{24.78}{83}{42.38}{27.21}{84}
\emline{42.38}{27.21}{85}{43.67}{31.67}{86}
\emline{46.33}{13.67}{87}{46.33}{4.33}{88}
\emline{58.00}{14.00}{89}{58.00}{4.67}{90}
\emline{71.67}{28.33}{91}{71.67}{18.67}{92}
\emline{71.67}{28.33}{93}{85.33}{4.67}{94}
\emline{72.00}{28.67}{95}{85.00}{18.67}{96}
\emline{69.67}{30.67}{97}{68.72}{28.35}{98}
\emline{68.72}{28.35}{99}{67.94}{26.04}{100}
\emline{67.94}{26.04}{101}{67.30}{23.73}{102}
\emline{67.30}{23.73}{103}{66.83}{21.42}{104}
\emline{66.83}{21.42}{105}{66.51}{19.11}{106}
\emline{66.51}{19.11}{107}{66.35}{16.80}{108}
\emline{66.35}{16.80}{109}{66.34}{14.48}{110}
\emline{66.34}{14.48}{111}{66.49}{12.17}{112}
\emline{66.49}{12.17}{113}{66.79}{9.86}{114}
\emline{66.79}{9.86}{115}{67.26}{7.55}{116}
\emline{67.26}{7.55}{117}{67.87}{5.24}{118}
\emline{67.87}{5.24}{119}{69.00}{2.00}{120}
\emline{71.33}{14.00}{121}{71.33}{4.67}{122}
\emline{71.33}{4.67}{123}{71.33}{4.67}{124}
\emline{85.33}{14.00}{125}{85.33}{5.00}{126}
\end{picture}
\special{em:linewidth 0.4pt} \unitlength 1.00mm
\linethickness{0.4pt}
\begin{picture}(119.00,36.00)
\put(3.33,3.33){\circle{5.33}} \put(16.67,3.33){\circle{5.33}}
\put(31.00,3.33){\circle{5.33}} \put(45.33,3.33){\circle{5.33}}
\put(58.67,3.33){\circle{5.33}} \put(74.00,3.33){\circle{5.33}}
\put(88.67,3.00){\circle{5.33}} \put(102.00,3.00){\circle{5.33}}
\put(116.33,3.00){\circle{5.33}} \put(17.33,19.00){\circle{5.33}}
\put(30.67,19.00){\circle{5.33}} \put(45.00,19.00){\circle{5.33}}
\put(59.33,19.00){\circle{5.33}} \put(72.67,19.00){\circle{5.33}}
\put(88.00,19.00){\circle{5.33}} \put(102.67,19.00){\circle{5.33}}
\put(110.00,33.33){\circle{5.33}} \put(80.00,33.33){\circle{5.33}}
\put(42.67,33.33){\circle{5.33}}
\put(42.67,33.33){\makebox(0,0)[cc]{$A_6$}}
\put(80.00,33.33){\makebox(0,0)[cc]{$A_7$}}
\put(110.00,33.67){\makebox(0,0)[cc]{$A_8$}}
\put(17.33,19.00){\makebox(0,0)[cc]{$D_7$}}
\put(30.67,19.00){\makebox(0,0)[cc]{$D_8$}}
\put(45.00,19.00){\makebox(0,0)[cc]{$D_9$}}
\put(59.33,19.00){\makebox(0,0)[cc]{$D_{10}$}}
\put(72.67,19.00){\makebox(0,0)[cc]{$D_{11}$}}
\put(88.00,19.00){\makebox(0,0)[cc]{$D_{12}$}}
\put(102.67,19.00){\makebox(0,0)[cc]{$D_{13}$}}
\put(3.00,3.00){\makebox(0,0)[cc]{$N_7$}}
\put(16.33,3.00){\makebox(0,0)[cc]{$N_8$}}
\put(31.00,3.00){\makebox(0,0)[cc]{$N_9$}}
\put(45.33,3.00){\makebox(0,0)[cc]{$N_{10}$}}
\put(58.67,3.33){\makebox(0,0)[cc]{$N_{11}$}}
\put(74.00,3.33){\makebox(0,0)[cc]{$N_{12}$}}
\put(88.67,3.00){\makebox(0,0)[cc]{$N_{13}$}}
\put(102.00,3.00){\makebox(0,0)[cc]{$N_{14}$}}
\put(116.33,3.00){\makebox(0,0)[cc]{$N_{15}$}}
\emline{42.67}{30.67}{1}{17.00}{21.33}{2}
\emline{42.33}{30.33}{3}{30.67}{21.33}{4}
\emline{42.67}{30.33}{5}{45.33}{21.67}{6}
\emline{42.67}{30.33}{7}{59.33}{21.67}{8}
\emline{42.67}{30.33}{9}{3.33}{6.00}{10}
\emline{30.67}{16.33}{11}{3.33}{5.67}{12}
\emline{42.67}{30.33}{13}{31.00}{6.00}{14}
\emline{42.67}{30.33}{15}{42.02}{27.78}{16}
\emline{42.02}{27.78}{17}{41.50}{25.25}{18}
\emline{41.50}{25.25}{19}{41.10}{22.75}{20}
\emline{41.10}{22.75}{21}{40.83}{20.26}{22}
\emline{40.83}{20.26}{23}{40.69}{17.80}{24}
\emline{40.69}{17.80}{25}{40.68}{15.36}{26}
\emline{40.68}{15.36}{27}{40.80}{12.94}{28}
\emline{40.80}{12.94}{29}{41.04}{10.54}{30}
\emline{41.04}{10.54}{31}{41.41}{8.16}{32}
\emline{41.41}{8.16}{33}{41.90}{5.80}{34}
\emline{41.90}{5.80}{35}{42.67}{3.00}{36}
\emline{42.33}{30.33}{37}{58.67}{5.67}{38}
\emline{17.00}{16.33}{39}{3.67}{6.33}{40}
\emline{42.33}{18.67}{41}{16.67}{6.00}{42}
\emline{45.00}{16.33}{43}{31.00}{6.00}{44}
\emline{59.33}{16.33}{45}{45.33}{6.00}{46}
\emline{58.67}{6.00}{47}{58.67}{16.00}{48}
\emline{77.33}{33.33}{49}{58.67}{6.00}{50}
\emline{79.67}{30.67}{51}{72.67}{21.67}{52}
\emline{79.67}{30.33}{53}{88.33}{21.67}{54}
\emline{79.67}{30.33}{55}{74.00}{6.00}{56}
\emline{80.00}{30.33}{57}{88.33}{5.67}{58}
\emline{79.67}{30.33}{59}{82.06}{29.05}{60}
\emline{82.06}{29.05}{61}{84.32}{27.70}{62}
\emline{84.32}{27.70}{63}{86.46}{26.29}{64}
\emline{86.46}{26.29}{65}{88.48}{24.82}{66}
\emline{88.48}{24.82}{67}{90.37}{23.28}{68}
\emline{90.37}{23.28}{69}{92.13}{21.67}{70}
\emline{92.13}{21.67}{71}{93.77}{20.01}{72}
\emline{93.77}{20.01}{73}{95.28}{18.27}{74}
\emline{95.28}{18.27}{75}{96.67}{16.48}{76}
\emline{96.67}{16.48}{77}{97.93}{14.62}{78}
\emline{97.93}{14.62}{79}{99.07}{12.69}{80}
\emline{99.07}{12.69}{81}{100.08}{10.71}{82}
\emline{100.08}{10.71}{83}{100.97}{8.65}{84}
\emline{100.97}{8.65}{85}{102.00}{5.67}{86}
\emline{88.00}{16.67}{87}{88.00}{5.67}{88}
\emline{102.67}{16.00}{89}{102.67}{5.67}{90}
\emline{102.67}{16.00}{91}{116.33}{5.67}{92}
\emline{102.67}{21.33}{93}{110.00}{30.67}{94}
\emline{110.00}{30.33}{95}{109.77}{28.05}{96}
\emline{109.77}{28.05}{97}{109.71}{25.76}{98}
\emline{109.71}{25.76}{99}{109.82}{23.48}{100}
\emline{109.82}{23.48}{101}{110.09}{21.20}{102}
\emline{110.09}{21.20}{103}{110.53}{18.91}{104}
\emline{110.53}{18.91}{105}{111.13}{16.63}{106}
\emline{111.13}{16.63}{107}{111.90}{14.35}{108}
\emline{111.90}{14.35}{109}{112.83}{12.06}{110}
\emline{112.83}{12.06}{111}{113.93}{9.78}{112}
\emline{113.93}{9.78}{113}{116.33}{5.67}{114}
\emline{80.00}{30.67}{115}{103.00}{21.33}{116}
\emline{58.67}{6.33}{117}{72.67}{16.33}{118}
\emline{72.67}{16.33}{119}{74.00}{6.00}{120}
\emline{16.67}{5.67}{121}{16.76}{7.43}{122}
\emline{16.76}{7.43}{123}{17.03}{9.16}{124}
\emline{17.03}{9.16}{125}{17.49}{10.85}{126}
\emline{17.49}{10.85}{127}{18.12}{12.51}{128}
\emline{18.12}{12.51}{129}{18.94}{14.12}{130}
\emline{18.94}{14.12}{131}{19.94}{15.70}{132}
\emline{19.94}{15.70}{133}{21.12}{17.24}{134}
\emline{21.12}{17.24}{135}{22.49}{18.74}{136}
\emline{22.49}{18.74}{137}{24.03}{20.21}{138}
\emline{24.03}{20.21}{139}{25.76}{21.63}{140}
\emline{25.76}{21.63}{141}{27.67}{23.02}{142}
\emline{27.67}{23.02}{143}{29.76}{24.37}{144}
\emline{29.76}{24.37}{145}{32.04}{25.69}{146}
\emline{32.04}{25.69}{147}{34.49}{26.96}{148}
\emline{34.49}{26.96}{149}{37.13}{28.20}{150}
\emline{37.13}{28.20}{151}{42.33}{30.33}{152}
\end{picture}
\special{em:linewidth 0.4pt} \unitlength 1.00mm
\linethickness{0.4pt}
\begin{picture}(104.67,36.34)
\put(3.33,3.33){\circle{5.33}} \put(16.67,3.33){\circle{5.33}}
\put(31.00,3.33){\circle{5.33}} \put(58.67,3.33){\circle{5.33}}
\put(74.00,3.33){\circle{5.33}} \put(88.67,3.00){\circle{5.33}}
\put(102.00,3.00){\circle{5.33}} \put(17.33,19.00){\circle{5.33}}
\put(30.67,19.00){\circle{5.33}} \put(59.33,19.00){\circle{5.33}}
\put(72.67,19.00){\circle{5.33}} \put(80.00,33.33){\circle{5.33}}
\put(3.33,19.00){\circle{5.33}} \put(10.00,33.00){\circle{5.33}}
\emline{10.00}{30.33}{1}{3.33}{21.67}{2}
\emline{10.00}{30.00}{3}{17.33}{21.67}{4}
\emline{3.00}{16.33}{5}{3.00}{6.00}{6}
\emline{16.67}{16.33}{7}{16.67}{6.00}{8}
\emline{10.00}{30.00}{9}{10.49}{27.70}{10}
\emline{10.49}{27.70}{11}{10.83}{25.39}{12}
\emline{10.83}{25.39}{13}{11.01}{23.08}{14}
\emline{11.01}{23.08}{15}{11.03}{20.77}{16}
\emline{11.03}{20.77}{17}{10.89}{18.45}{18}
\emline{10.89}{18.45}{19}{10.59}{16.12}{20}
\emline{10.59}{16.12}{21}{10.14}{13.79}{22}
\emline{10.14}{13.79}{23}{9.52}{11.45}{24}
\emline{9.52}{11.45}{25}{8.75}{9.11}{26}
\emline{8.75}{9.11}{27}{7.82}{6.76}{28}
\emline{7.82}{6.76}{29}{6.00}{3.00}{30}
\put(40.33,33.33){\circle{5.33}} \put(46.00,18.67){\circle{5.33}}
\put(46.00,3.00){\circle{5.33}}
\emline{30.33}{21.67}{31}{40.33}{30.67}{32}
\emline{40.00}{30.67}{33}{30.67}{6.00}{34}
\emline{40.00}{30.67}{35}{39.83}{27.99}{36}
\emline{39.83}{27.99}{37}{39.79}{25.37}{38}
\emline{39.79}{25.37}{39}{39.90}{22.83}{40}
\emline{39.90}{22.83}{41}{40.16}{20.35}{42}
\emline{40.16}{20.35}{43}{40.55}{17.94}{44}
\emline{40.55}{17.94}{45}{41.09}{15.60}{46}
\emline{41.09}{15.60}{47}{41.78}{13.33}{48}
\emline{41.78}{13.33}{49}{42.60}{11.13}{50}
\emline{42.60}{11.13}{51}{43.57}{9.00}{52}
\emline{43.57}{9.00}{53}{45.67}{5.33}{54}
\emline{30.33}{16.33}{55}{30.33}{6.00}{56}
\emline{33.33}{19.00}{57}{45.33}{5.33}{58}
\emline{46.00}{16.00}{59}{46.00}{5.67}{60}
\put(59.33,33.67){\circle{5.33}}
\emline{59.33}{31.00}{61}{45.67}{21.33}{62}
\emline{59.33}{30.67}{63}{46.00}{5.67}{64}
\emline{59.33}{30.33}{65}{60.68}{28.38}{66}
\emline{60.68}{28.38}{67}{61.82}{26.39}{68}
\emline{61.82}{26.39}{69}{62.75}{24.37}{70}
\emline{62.75}{24.37}{71}{63.47}{22.31}{72}
\emline{63.47}{22.31}{73}{63.99}{20.22}{74}
\emline{63.99}{20.22}{75}{64.29}{18.09}{76}
\emline{64.29}{18.09}{77}{64.39}{15.93}{78}
\emline{64.39}{15.93}{79}{64.28}{13.74}{80}
\emline{64.28}{13.74}{81}{63.97}{11.51}{82}
\emline{63.97}{11.51}{83}{63.44}{9.25}{84}
\emline{63.44}{9.25}{85}{62.71}{6.95}{86}
\emline{62.71}{6.95}{87}{61.33}{3.67}{88}
\emline{59.00}{30.67}{89}{74.00}{6.00}{90}
\emline{62.00}{33.33}{91}{64.74}{32.16}{92}
\emline{64.74}{32.16}{93}{67.33}{30.95}{94}
\emline{67.33}{30.95}{95}{69.78}{29.69}{96}
\emline{69.78}{29.69}{97}{72.07}{28.39}{98}
\emline{72.07}{28.39}{99}{74.22}{27.05}{100}
\emline{74.22}{27.05}{101}{76.22}{25.67}{102}
\emline{76.22}{25.67}{103}{78.07}{24.24}{104}
\emline{78.07}{24.24}{105}{79.78}{22.77}{106}
\emline{79.78}{22.77}{107}{81.33}{21.25}{108}
\emline{81.33}{21.25}{109}{82.74}{19.69}{110}
\emline{82.74}{19.69}{111}{84.00}{18.09}{112}
\emline{84.00}{18.09}{113}{85.11}{16.45}{114}
\emline{85.11}{16.45}{115}{86.07}{14.76}{116}
\emline{86.07}{14.76}{117}{86.89}{13.03}{118}
\emline{86.89}{13.03}{119}{87.56}{11.25}{120}
\emline{87.56}{11.25}{121}{88.08}{9.43}{122}
\emline{88.08}{9.43}{123}{88.45}{7.57}{124}
\emline{88.45}{7.57}{125}{88.67}{5.67}{126}
\emline{59.00}{30.67}{127}{59.00}{21.67}{128}
\emline{59.00}{31.00}{129}{72.67}{21.67}{130}
\emline{46.00}{16.00}{131}{59.00}{6.00}{132}
\emline{59.00}{16.33}{133}{59.00}{6.00}{134}
\emline{59.00}{16.67}{135}{74.00}{6.33}{136}
\emline{80.00}{30.67}{137}{72.67}{21.67}{138}
\put(9.67,32.67){\makebox(0,0)[cc]{$A_9$}}
\put(40.33,33.33){\makebox(0,0)[cc]{$A_{10}$}}
\put(59.33,33.33){\makebox(0,0)[cc]{$A_{11}$}}
\put(80.00,33.00){\makebox(0,0)[cc]{$A_{12}$}}
\put(3.33,18.67){\makebox(0,0)[cc]{$D_{14}$}}
\put(17.33,18.67){\makebox(0,0)[cc]{$D_{15}$}}
\put(30.33,19.00){\makebox(0,0)[cc]{$D_{16}$}}
\put(46.00,19.00){\makebox(0,0)[cc]{$D_{17}$}}
\put(59.33,19.00){\makebox(0,0)[cc]{$D_{18}$}}
\put(72.67,19.00){\makebox(0,0)[cc]{$D_{19}$}}
\put(3.33,3.67){\makebox(0,0)[cc]{$N_{16}$}}
\put(16.67,3.67){\makebox(0,0)[cc]{$N_{17}$}}
\put(31.00,3.67){\makebox(0,0)[cc]{$N_{18}$}}
\put(46.00,3.00){\makebox(0,0)[cc]{$N_{19}$}}
\put(58.67,3.33){\makebox(0,0)[cc]{$N_{20}$}}
\put(74.00,3.33){\makebox(0,0)[cc]{$N_{21}$}}
\put(88.67,2.67){\makebox(0,0)[cc]{$N_{22}$}}
\put(102.00,3.00){\makebox(0,0)[cc]{$N_{23}$}}
\emline{82.67}{33.33}{139}{85.07}{32.08}{140}
\emline{85.07}{32.08}{141}{87.31}{30.77}{142}
\emline{87.31}{30.77}{143}{89.39}{29.40}{144}
\emline{89.39}{29.40}{145}{91.32}{27.98}{146}
\emline{91.32}{27.98}{147}{93.08}{26.49}{148}
\emline{93.08}{26.49}{149}{94.69}{24.94}{150}
\emline{94.69}{24.94}{151}{96.14}{23.34}{152}
\emline{96.14}{23.34}{153}{97.43}{21.67}{154}
\emline{97.43}{21.67}{155}{98.56}{19.95}{156}
\emline{98.56}{19.95}{157}{99.53}{18.16}{158}
\emline{99.53}{18.16}{159}{100.35}{16.32}{160}
\emline{100.35}{16.32}{161}{101.00}{14.41}{162}
\emline{101.00}{14.41}{163}{101.50}{12.45}{164}
\emline{101.50}{12.45}{165}{101.84}{10.43}{166}
\emline{101.84}{10.43}{167}{102.02}{8.34}{168}
\emline{102.02}{8.34}{169}{102.00}{5.33}{170}
\emline{75.33}{19.00}{171}{88.67}{5.67}{172}
\emline{75.33}{18.67}{173}{102.00}{5.67}{174}
\emline{16.67}{6.00}{175}{18.10}{7.77}{176}
\emline{18.10}{7.77}{177}{19.29}{9.49}{178}
\emline{19.29}{9.49}{179}{20.26}{11.18}{180}
\emline{20.26}{11.18}{181}{20.99}{12.82}{182}
\emline{20.99}{12.82}{183}{21.48}{14.43}{184}
\emline{21.48}{14.43}{185}{21.74}{15.99}{186}
\emline{21.74}{15.99}{187}{21.77}{17.52}{188}
\emline{21.77}{17.52}{189}{21.56}{19.00}{190}
\emline{21.56}{19.00}{191}{21.12}{20.45}{192}
\emline{21.12}{20.45}{193}{20.45}{21.85}{194}
\emline{20.45}{21.85}{195}{19.54}{23.21}{196}
\emline{19.54}{23.21}{197}{18.40}{24.54}{198}
\emline{18.40}{24.54}{199}{17.03}{25.82}{200}
\emline{17.03}{25.82}{201}{15.42}{27.06}{202}
\emline{15.42}{27.06}{203}{13.57}{28.27}{204}
\emline{13.57}{28.27}{205}{9.67}{30.33}{206}
\end{picture}
\end{center}
\caption{\label{f50} The interval graph corresponding to the
intervals of Figure \ref{f40}. }
\end{figure}
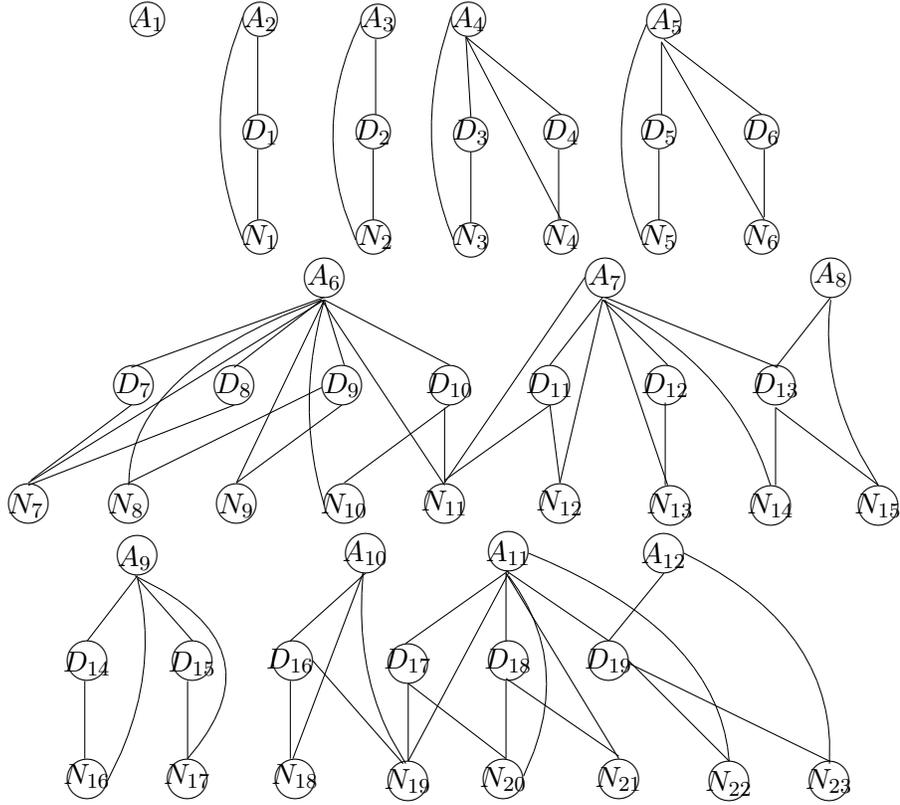
 }

For the above example {it is observed} that the interval graph is
disconnected and has eight components. But, if {the} large number
of channels {are considered} then the graph {may be} connected. If
the graph becomes disconnected then apply the algorithm MWKF to
each component and combine the solutions obtained from all
components. The combined solution is the final solution of the
problem.


\subsection{Computational result}
\ccc{The algorithm is implemented in C and the results for the
problem consider in this section are given below. The weights
corresponding to the programmes are taken randomly between 1 and
9.}

\medskip\noindent\ccc{\textit{Step 1:} All maximal cliques are \\
$c_1=(1), c_2=(2, 3, 4), c_3=(5, 6, 7), c_4=(8, 9,10), c_5=(10,
11, 12), c_6=(13, 14, 15),$\\ $c_7=(15, 16, 17),$ $ c_8=(18, 20,
25), c_9=(19, 20, 25), c_{10}=(21, 22, 25), c_{11}=(22, 23,
25),$\\
$c_{12}=(24, 25, 26)$, $c_{13}=(25, 26, 27),$ $c_{14}=(27, 29,
32), c_{15}=(28, 29, 32),$\\ $c_{16}=(30, 31, 32), c_{17}=(32, 33,
34), c_{18}=(34, 35, 36), c_{19}=(37, 38, 41),$\\ $c_{20}=(39, 40,
41), c_{21}=(42, 43, 44), c_{22}=(43, 44, 45), c_{23}=(45, 46),$\\
$c_{24}=(46, 47, 50), c_{25}=(47, 48, 50), c_{26}=(48, 49, 50),
c_{27}=(50, 51, 52), c_{28}=(52, 53, 54)$. }

\medskip\noindent\ccc{\textit{Step 2:} The nodes of $N$ are\\
$C_0, C_1, C_2, C_3, C_4, C_5, C_6, C_7, C_8, C_9, C_{10}, C_{11},
C_{12}, C_{13}, C_{14},$\\ $C_{15}, C_{16}, C_{17}, C_{18},
C_{19}, C_{20}, C_{21}, C_{22}, C_{23}, C_{24}, C_{25}, C_{26},
C_{27}, C_{28}$.}

\medskip\noindent\ccc{$i$-arcs: $e_1: (C_0,C_1),$  $e_2:
(C_1,C_2),$ $e_3: (C_1,C_2),$ $e_4: (C_1,C_2),$ $e_5: (C_2,C_3),$
$e_6: (C_2,C_3),$ $e_7: (C_2,C_3),$ $e_8: (C_3,C_4),$ $e_9:
(C_3,C_4),$
$e_{10}: (C_3,C_5),$ $e_{11}: (C_4,C_5),$ $e_{12}: (C_4,C_5),$\\
$e_{13}: (C_5,C_6),$ $e_{14}: (C_5,C_6),$ $e_{15}: (C_5,C_7),$
$e_{16}: (C_6,C_7),$ $e_{17}: (C_6,C_7),$ $e_{18}: (C_7,C_8),$\\
$e_{19}: (C_8,C_9),$ $e_{20}: (C_7,C_9),$ $e_{21}: (C_9,C_{10}),$
$e_{22}: (C_9,C_{11}),$ $e_{23}: (C_{10},C_{11}),$ $e_{24}:
(C_{11},C_{12}),$\\ $e_{25}: (C_{7},C_{13}),$ $e_{26}:
(C_{11},C_{13}),$ $e_{27}: (C_{12},C_{14}),$ $e_{28}:
(C_{14},C_{15}),$ $e_{29}: (C_{13},C_{15}),$ $e_{30}:
(C_{15},C_{16}),$ $e_{31}: (C_{15},C_{16}),$ $e_{32}:
(C_{13},C_{17}),$ $e_{33}: (C_{16},C_{17}),$ $e_{34}:
(C_{16},C_{18}),$ $e_{35}: (C_{17},C_{18}),$ $e_{36}:
(C_{17},C_{18}),$ $e_{37}: (C_{18},C_{19}),$ $e_{38}:
(C_{18},C_{19}),$ $e_{39}: (C_{19},C_{20}),$ $e_{40}:
(C_{19},C_{20}),$ $e_{41}: (C_{18},C_{20}),$ $e_{42}:
(C_{20},C_{21}),$ $e_{43}: (C_{20},C_{22}),$ $e_{44}:
(C_{20},C_{22}),$ $e_{45}: (C_{21},C_{23}),$ $e_{46}:
(C_{22},C_{24}),$ $e_{47}: (C_{23},C_{25}),$ $e_{48}:
(C_{24},C_{26}),$ $e_{49}: (C_{25},C_{26}),$ $e_{50}:
(C_{23},C_{27}),$ $e_{51}: (C_{26},C_{27}),$ $e_{52}:
(C_{26},C_{28}),$ $e_{53}: (C_{27},C_{28}),$ $e_{54}:
(C_{27},C_{28}).$}

\medskip\noindent\ccc{$c$-arcs: $e_{54+i}: (C_{i-1}, C_i), i=1, 2,
\ldots, 28$.}

\medskip\noindent\ccc{\textit{Step 3:} $\pi$ values of the nodes:\\
$\pi(C_0)=110$, $\pi(C_1)=107$, $\pi(C_2)=102$, $\pi(C_3)=95$,
$\pi(C_4)=87$, $\pi(C_5)=84$, $\pi(C_6)=80$, $\pi(C_7)=75$,
$\pi(C_8)=71$, $\pi(C_9)=64$, $\pi(C_{10})=57$, $\pi(C_{11})=49$,
$\pi(C_{12})=47$, $\pi(C_{13})=42$, $\pi(C_{14})=43$,
$\pi(C_{15})=38$, $\pi(C_{16})=32$, $\pi(C_{17})=28$,
$\pi(C_{18})=25$, $\pi(C_{19})=21$, $\pi(C_{20})=19$,
$\pi(C_{21})=16$, $\pi(C_{22})=10$, $\pi(C_{23})=10$,
$\pi(C_{24})=9$, $\pi(C_{25})=9$, $\pi(C_{26})=5$,
$\pi(C_{27})=2$, $\pi(C_{28})=0$.}

\medskip
\noindent\ccc{\textit{Step 4:} The weights of the arcs of the
network $N^U$.}

\medskip\noindent\ccc{$i$-arcs: $e_1=0$, $e_2=4$, $e_3=0$, $e_4=0$,
$e_5=5$, $e_6=0$, $e_7=0$, $e_8=0$, $e_9=0$, $e_{10}=8$,
$e_{11}=0$, $e_{12}=0$, $e_{13}=0$, $e_{14}=0$, $e_{15}=7$,
$e_{16}=0$, $e_{17}=0$, $e_{18}=0$, $e_{19}=0$, $e_{20}=7$,
$e_{21}=0$, $e_{22}=7$, $e_{23}=0$, $e_{24}=0$, $e_{25}=28$,
$e_{26}=5$, $e_{27}=0$, $e_{28}=0$, $e_{29}=0$, $e_{30}=1$,
$e_{31}=0$, $e_{32}=10$, $e_{33}=0$, $e_{34}=1$, $e_{35}=0$,
$e_{36}=0$, $e_{37}=0$, $e_{38}=3$, $e_{39}=0$, $e_{40}=0$,
$e_{41}=2$, $e_{42}=0$, $e_{43}=6$, $e_{44}=7$, $e_{45}=0$,
$e_{46}=0$, $e_{47}=0$, $e_{48}=0$, $e_{49}=0$, $e_{50}=4$,
$e_{51}=0$,$e_{52}=3$, $e_{53}=1$, $e_{54}=0$.}

\medskip\noindent\ccc{$c$-arcs: $e_{55}=3$, $e_{56}=5$, $e_{57}=7$,
$e_{58}=8$, $e_{59}=3$, $e_{60}=4$, $e_{61}=5$, $e_{62}=4$,
$e_{63}=7$, $e_{64}=7$, $e_{65}=8$, $e_{66}=2$, $e_{67}=5$,
$e_{68}=1$, $e_{69}=5$, $e_{70}=6$, $e_{71}=4$, $e_{72}=3$,
$e_{73}=4$, $e_{74}=2$, $e_{75}=3$, $e_{76}=6$, $e_{77}=0$,
$e_{78}=1$, $e_{79}=0$, $e_{80}=4$, $e_{81}=3$, $e_{82}=2$.}

\medskip\noindent\ccc{\textit{Step 5:} Here we consider $k=2$. The two
paths $P_1$ and $P_2$ are given below.\\
$P_1:
C_0-e_1-C_{1}-e_3-C_{2}-e_6-C_{3}-e_8-C_{4}-e_{11}-C_{5}-e_{13}-C_{6}
-e_{16}-C_{7}-e_{18}-C_{8}-e_{19}-C_{9}-e_{21}-C_{10}-e_{23}-C_{11}-e_{24}
-C_{12}-e_{27}-C_{14}-e_{28}-C_{15}-e_{31}-C_{16}-e_{33}-C_{17}-e_{35}-C_{18}
-e_{37}-C_{19}-e_{40}-C_{20}-e_{42}-C_{21}-e_{45}-C_{23}-e_{47}-C_{25}-
e_{49}-C_{26}-e_{51}-C_{27}-e_{52}-C_{28}$.}

\medskip\noindent \ccc{$P_2:
C_0-e_{55}-C_{1}-e_4-C_{2}-e_7-C_{3}-e_9-C_{4}-e_{12}-C_{5}-e_{14}-C_{6}
-e_{17}-C_{7}-e_{20}-C_{9}-e_{22}-C_{11}-e_{26}-C_{13}-e_{29}-C_{15}-e_{30}
-C_{16}-e_{34}-C_{18}-e_{38}-C_{19}-e_{39}-C_{20}-e_{43}-C_{22}-e_{46}-C_{24}
-e_{48}-C_{26}-e_{52}-C_{28}$.}

\medskip\noindent\ccc{\textit{Step 6 and Step 7:}
The set of $i$-arcs on the path $P_1$ is\\
$X_1=\{e_1,e_3,e_6,e_8,e_{11},e_{13},e_{16},e_{18},e_{19},e_{21},e_{23},e_{24},
e_{27},e_{28},$\\ \tab\tab $e_{31},e_{33},e_{35},e_{37},e_{40},e_{42},e_{45},e_{47},e_{49},e_{51},e_{52}\}$\\
and for the path $P_2$ is\\
$X_1=\{e_4,e_7,e_9,e_{12},e_{14},e_{17},e_{20},e_{22},e_{26},e_{29},e_{30},
e_{34},e_{38},e_{39},e_{43},e_{46},e_{48},e_{52}\}$.}

\ccc{The set of vertices corresponding to the $i$-arcs of $X_1$
and $X_2$ are \\
$H_1=\{A_1, N_1, N_2, N_3, N_4, N_5, N_6, D_7, D_8, N_8, N_9,
N_{10},
N_{11}, N_{12}, N_{13}, N_{14}, N_{15}, D_{14},$\\
\tab\tab $D_{15}, N_{18}, N_{19}, N_{20}, N_{21}, N_{22}, A_{12}\}$\\
and $H_2=\{D_1, D_2, D_3, D_4, D_5, D_6, N_7, D_9, D_{10}, D_{11},
D_{12}, D_{13}, N_{16}, N_{17}, D_{16}, D_{17}, D_{18}, D_{19}\}$.
} 

 The weights of $H_1$ and $H_2$ are respectively 110 and 74 and
the total weight of 2-colour set is 184.

Another 2-colour set of this problem is given below.\\
$H_1=\{N_1, N_2, N_3, N_4, N_5, N_6, N_7, N_8, N_9, N_{10},
N_{11}, N_{12}, N_{13}, N_{14}, N_{15}, N_{16}, N_{17}, N_{18}, N_{19}, $\\
\tab\tab $N_{20}, N_{21}, N_{22}, A_{12}\}$\\
$H_2=\{A_1, D_1, D_2, D_3, D_4, D_5, D_6, D_7, D_8, D_9, D_{10},
D_{11}, D_{12}, D_{13}, D_{14}, D_{15}, D_{16}, D_{17}, D_{18},
D_{19}\}$.

In this solution the weights of $H_1$ and $H_2$ are respectively
97 and 87 and the total weight of 2-colour set is also 184.

Note that the weight of two 2-colour sets are same and this is
equal to {\bf
184}. Also, in both the cases\\
(i) $H_1\cap H_2=\phi$,\\
(ii) $H_1\cup H_2\subset V$,\\
(iii) Each colour set $H_1$ and $H_2$ forms an independent set,\\
(iv) Weight of $H_1\cup H_2$ is maximum among all other 2-colour
sets.

The following result follows form Theorem \ref{t40}.
\begin{theorem}
The {programme slots} for $k$ parallel sessions with maximum
number of viewers can be selected using $O(kMn^2)$ time, where $n$
is the total number of programmes in all channels and $M$ is the
upper bound of viewers of all programmes of all channels.
\end{theorem}

\section{Conclusion}
In this paper, a method {is described} for the ACs to display
their advertisement in {$k$ parallel} sessions in different
television channels. The objective of the ACs are to catch the
maximum number of viewers. The companies want to attract more and
more viewers through their advertisement. This real life problem
is modelled as an interval graph. The various programme slots are
considered as interval. The method we have developed has been
solved by converting the problem into maximum weight $k$-colouring
problem on interval graph. The number of viewers of a particular
programme slot is taken as the weight of that corresponding
interval. While solving the problem we have not considered the
subscription rate of the programmes of various channels. We are
trying to solve this problem {by taking into} account the
subscription rate as another objective.

It is obvious, that the companies try to select such programme
slots whose number of viewers are very high. Sometimes, it may
also happen that the subscription rate of a particular programme
having large number of viewers is very high. It becomes difficult
for a company to {afford} this high price. In this case, the
company may discard this high price programme slot using our
proposed algorithm by modifying the weight of the corresponding
programme slot or interval by assigning the weight to zero.

\end{document}